%% file: main.tex
\documentclass[webpdf,contemporary,large]{oup-authoring-template}

\graphicspath{{Fig/}}
\usepackage{float}
\definecolor{tablegroupgray}{gray}{0.80}

\theoremstyle{thmstyleone}

\theoremstyle{thmstyletwo}

\theoremstyle{thmstylethree}

\newcommand{\best}[1]{\textbf{\boldmath #1}}

\begin{document}

\journaltitle{Bioinformatics}
\DOI{DOI added during production}
\copyrightyear{2026}
\pubyear{2026}
\vol{XX}
\issue{x}
\access{Advance Access Publication Date: to be assigned}
\appnotes{Original Paper}
\firstpage{1}

\title[DMT-Dens for Biological Visualization]{DMT-Dens: Density-preserving manifold visualization for biological data}

\author[1]{Ruizhe Wang}
\author[1]{Yixuan Dong}
\author[1]{Bolin Yang}
\author[1]{Bingo Wing-Kuen Ling}
\author[1]{Fuji Yang}
\author[1,$\ast$]{Zelin Zang}

\address[1]{\orgname{Tsientang Institute for Advanced Study}, \orgaddress{\state{Hangzhou, Zhejiang}, \country{China}}}

\corresp[$\ast$]{Corresponding author. \href{mailto:zangzelin@gmail.com}{zangzelin@gmail.com}}

\received{Date}{0}{2026}
\revised{Date}{0}{2026}
\accepted{Date}{0}{2026}

\abstract{\textbf{Motivation:} Low-dimensional embeddings are widely used to explore cell-state heterogeneity in single-cell and other high-dimensional biological data. Although many methods preserve local neighborhoods, they may distort the apparent sampling density of processed observations, altering the visual contrast between dense and sparse regions and complicating the interpretation of rare, transitional, or continuous cell-state populations.\\
\textbf{Results:} We present DMT-Dens, a parametric manifold-visualization method built on a latent-token Transformer encoder. The model integrates rank-based manifold alignment with hard-pair aggregation. To preserve density, it optimizes a loss based on the Pearson correlation between $k$-nearest-neighbor log-radius estimates in the processed input and two-dimensional embedding spaces. Benchmark evaluations demonstrate strong density preservation, particularly on biological datasets, while retaining competitive label separability.\\
\textbf{Availability:} Source code, data-processing scripts, and resolved experiment configurations are available at \url{https://github.com/Ruizhe-wang/DMT-Dens}.\\
\textbf{Contact:} \href{mailto:zangzelin@gmail.com}{zangzelin@gmail.com}\\
\textbf{Supplementary information:} Supplementary data are included after the main article in this arXiv version.}

\boxedtext{Key Messages}{
\begin{itemize}
\item DMT-Dens targets biological visualization settings in which density fidelity and label separability are evaluated together.
\item The method combines a parametric mapping and rank-based manifold alignment with an explicit kNN density-preservation loss.
\item The evaluation focuses on density correlation and local density correlation; SVC accuracy is reported as a descriptive label-separability diagnostic, with neighborhood and trajectory diagnostics provided as Supplementary Material.
\end{itemize}}

\maketitle

\input{sections/introduction}
\input{sections/methods}
\input{sections/results}
\input{sections/discussion}
\raggedbottom
\raggedend
\input{sections/backmatter}

\clearpage
\bibliographystyle{oup-abbrvnat}
\bibliography{topobranch_refs}

\clearpage
\onecolumn
\raggedbottom

\begin{center}
  \vspace*{6pt}
  {\LARGE\bfseries Supplementary Material\par}
  \vspace{6pt}
  {\large DMT-Dens: Density-preserving manifold visualization for biological data\par}
\end{center}
\vspace{12pt}

\setcounter{section}{0}
\setcounter{subsection}{0}
\setcounter{table}{0}
\setcounter{figure}{0}
\setcounter{equation}{0}
\setcounter{secnumdepth}{2}
\renewcommand{\thesection}{S\arabic{section}}
\renewcommand{\thesubsection}{S\arabic{section}.\arabic{subsection}}
\renewcommand{\thetable}{S\arabic{table}}
\renewcommand{\thefigure}{S\arabic{figure}}
\renewcommand{\theequation}{S\arabic{equation}}
\renewcommand{\theHsection}{supp.section.\arabic{section}}
\renewcommand{\theHsubsection}{supp.subsection.\arabic{section}.\arabic{subsection}}
\renewcommand{\theHtable}{supp.table.\arabic{table}}
\renewcommand{\theHfigure}{supp.figure.\arabic{figure}}
\renewcommand{\theHequation}{supp.equation.\arabic{equation}}

\input{sections/supplementary}

\end{document}

%% file: sections/introduction.tex
\section{Introduction}

\begin{figure*}[!t]
\centering
\includegraphics[width=\textwidth]{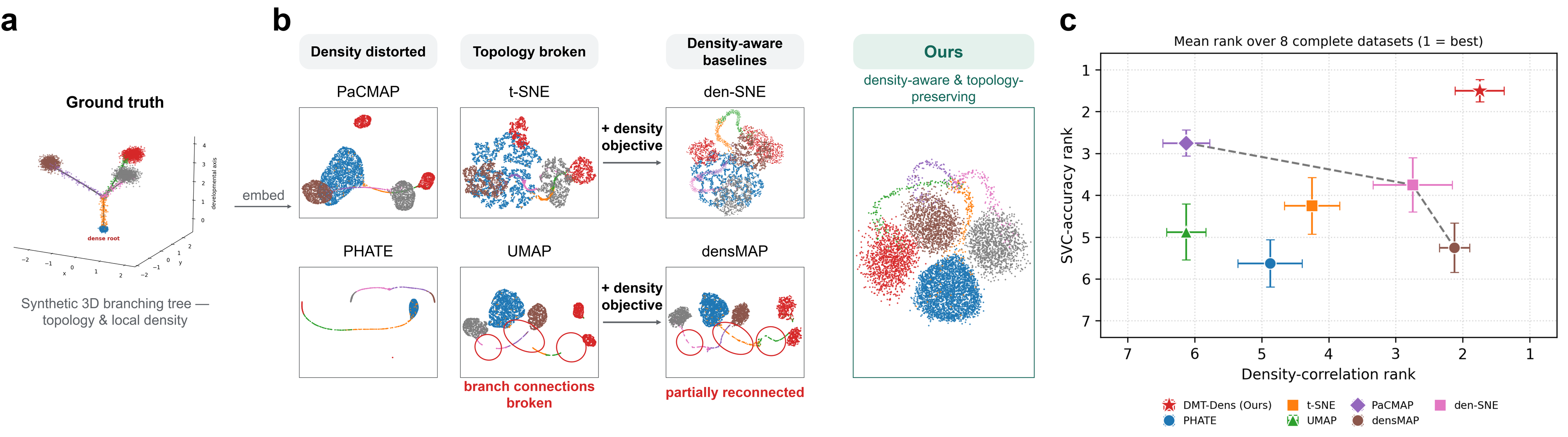}
\caption{Motivation for density-preserving biological visualization. \textbf{a,} Ground-truth synthetic three-dimensional branching tree with a dense root, sparse transitions, and multiple branch points, colored by branch identity. \textbf{b,} Two-dimensional embeddings of the same data. The displayed baseline embeddings differ in local density, apparent continuity, and adjacency. Adding a density objective to t-SNE and UMAP yields the density-aware baselines den-SNE and densMAP. The red outlines mark apparent disconnections in this representative UMAP embedding and their partial reconnection in densMAP. In the displayed DMT-Dens embedding, density variation and continuous paths are visually retained. \textbf{c,} Density preservation versus label separability across the benchmark datasets. Each point shows the mean rank of one method over the eight datasets with complete baseline coverage (density-correlation rank versus SVC-accuracy rank), and whiskers show the standard error across datasets. The dashed curve marks the Pareto frontier of the baseline methods.\label{fig:motivation}}
\end{figure*}

Low-dimensional visualization is widely used to explore high-dimensional data. In single-cell RNA sequencing (scRNA-seq), two-dimensional embeddings support cell-type annotation, quality control, and exploratory analysis of developmental continua and heterogeneous cell states \citep{becht2019umap,kobak2019art}. Deep manifold-learning methods have also been developed for structure-preserving visualization and data integration in single-cell and spatial transcriptomics \citep{xu2023dmt,xu2025poincaredmt,zang2025must}. These methods map sparse, noisy gene-expression profiles from high-dimensional space to two or three dimensions for visual inspection.

Common visualization methods do not preserve every property relevant to biological interpretation. Principal Component Analysis (PCA) provides a linear view of global variation. Nonlinear neighbor-embedding methods such as t-Distributed Stochastic Neighbor Embedding (t-SNE) and Uniform Manifold Approximation and Projection (UMAP) instead emphasize local structure and often produce clearly separated groups \citep{maaten2008visualizing,mcinnes2018umap,becht2019umap}. However, cluster area and spread in these embeddings do not reliably reflect cell-state abundance or transcriptional variability \citep{nguyen2019diffusion,chari2023specious,kobak2019art}. t-SNE may expand densely sampled regions and contract sparse parts of continuous trajectories \citep{narayan2021assessing,kobak2019art}, whereas UMAP may exaggerate population separation and alter local density \citep{chari2023specious,kobak2021initialization}. Here, density refers specifically to the sample density of the processed input representation. Experimental sampling determines which cells enter that representation, and tissue dissociation can induce transcriptional changes before density is calculated \citep{vandenbrink2017single}. Feature selection, normalization, and dimensionality-reduction steps such as PCA further define the geometry in which density is measured. Although processed-input density is not a calibrated measure of biological abundance, it is an empirical structural property presented to the visualization method. Preserving this relative density prevents the projection from introducing additional distortion and allows concentration and dispersion in the embedding to reflect the analyzed representation more faithfully \citep{narayan2021assessing}.

den-SNE and densMAP address density distortion by adding density-preservation terms to the t-SNE and UMAP objectives, respectively \citep{narayan2021assessing}. These methods improve density fidelity, but remain non-parametric and inherit the neighborhood-affinity objectives of their parent methods: pairwise probabilistic affinities in t-SNE/den-SNE and a fuzzy neighbor graph in UMAP/densMAP. Consequently, manifold organization is still governed by the corresponding parent attraction--repulsion formulation, with density preservation added to the inherited structural objective. This leaves a more specific methodological gap: an explicit parametric model that jointly learns neighborhood-rank structure and the relative density profile of the processed observations. Such joint optimization is relevant because biological data are often modeled as lying on a low-dimensional, nonlinear manifold whose observation density is nonuniform \citep{bunne2023cellot,biondo2025intrinsic,narayan2021assessing}. Within this scope, faithful sample-density visualization provides a controlled basis for examining heterogeneous and transitional regions while distinguishing projection-induced distortion from variation already present in the analyzed data.

We developed DMT-Dens to address this gap. Its primary contribution is an explicit density loss that aligns $k$-nearest-neighbor log-radius profiles between the processed input and the two-dimensional embedding. This loss is optimized within a parametric deep manifold-transformation model \citep{xu2023dmt}. The accompanying manifold objective represents neighborhood relations by rank affinity and emphasizes hard pairs during optimization. These mechanisms support neighborhood organization while the density term targets relative concentration and dispersion. Bidirectional affinity construction and two-scale density aggregation are additional refinements of the full objective.

To determine whether DMT-Dens improves density fidelity while retaining useful label and neighborhood structure, we compare its two-dimensional representations with those produced by t-SNE \citep{maaten2008visualizing}, UMAP \citep{mcinnes2018umap}, den-SNE and densMAP \citep{narayan2021assessing}, PaCMAP \citep{wang2021pacmap}, and PHATE \citep{moon2019phate}. The evaluation metrics, baseline settings, and hyperparameter-search protocol are described in the Methods and Supplementary Material. Under this common protocol, DMT-Dens achieves the highest density correlation on all four biological datasets and on six of the nine datasets overall, while its linear SVC accuracy ranks among the top two methods on seven datasets. Ablation experiments identify the density loss as the main source of improved density fidelity. Supplementary analyses of synthetic dyngen trajectories \citep{cannoodt2021dyngen} and a \emph{Caenorhabditis elegans} developmental time course \citep{packer2019celegan} provide an exploratory assessment of topology- and time-related diagnostics.

%% file: sections/methods.tex
\section{Material and methods}

\subsection{Problem definition}
Let $\mathbf{X}=[\mathbf{x}_1^\top,\dots,\mathbf{x}_N^\top]^\top\in\mathbb{R}^{N\times D}$ denote an input matrix with $N$ observations and $D$ features, where $\mathbf{x}_i\in\mathbb{R}^{D}$ is observation $i$. For single-cell data, an observation is a cell represented by processed gene-expression measurements or a derived feature vector. The formulation also covers the image, text, and sensor data used in our experiments. The encoder $e_\phi:\mathbb{R}^{D}\rightarrow\mathbb{R}^{40}$ maps $\mathbf{x}_i$ to a 40-dimensional latent representation $\mathbf{z}_i=e_\phi(\mathbf{x}_i)$. The projection module $p_\psi:\mathbb{R}^{40}\rightarrow\mathbb{R}^{2}$ maps $\mathbf{z}_i$ to a two-dimensional coordinate $\mathbf{y}_i=p_\psi(\mathbf{z}_i)$. We denote the full mapping by $f_\theta=p_\psi\circ e_\phi$, with $\theta=(\phi,\psi)$, and collect the coordinates in $\mathbf{Y}=[\mathbf{y}_1^\top,\dots,\mathbf{y}_N^\top]^\top\in\mathbb{R}^{N\times 2}$.

We optimize the map toward two objectives. (i) \emph{Manifold preservation}: the manifold-alignment loss encourages the embedding to retain relative neighborhood orderings from feature space. If $\mathbf{x}_i$ is closer to $\mathbf{x}_j$ than to $\mathbf{x}_k$, the loss favors the corresponding ordering between $\mathbf{y}_i$, $\mathbf{y}_j$, and $\mathbf{y}_k$. (ii) \emph{Density preservation}: the density loss encourages agreement between local-density profiles in feature and embedding spaces, helping to maintain the visual contrast between dense and sparse regions.

We express these objectives through a manifold-preservation loss $\mathcal{L}_{\mathrm{manifold}}$ and a density-preservation loss $\mathcal{L}_{\mathrm{dens}}$. The model parameters are estimated by minimizing their weighted sum,
\begin{equation}
\mathcal{L}(\theta)=\mathcal{L}_{\mathrm{manifold}}(\theta)+\lambda_d\,\mathcal{L}_{\mathrm{dens}}(\theta),
\label{eq:total-loss-overview}
\end{equation}
where $\lambda_d\ge 0$ controls the contribution of the density term.

\begin{figure*}[!t]
\centering
\includegraphics[width=\textwidth]{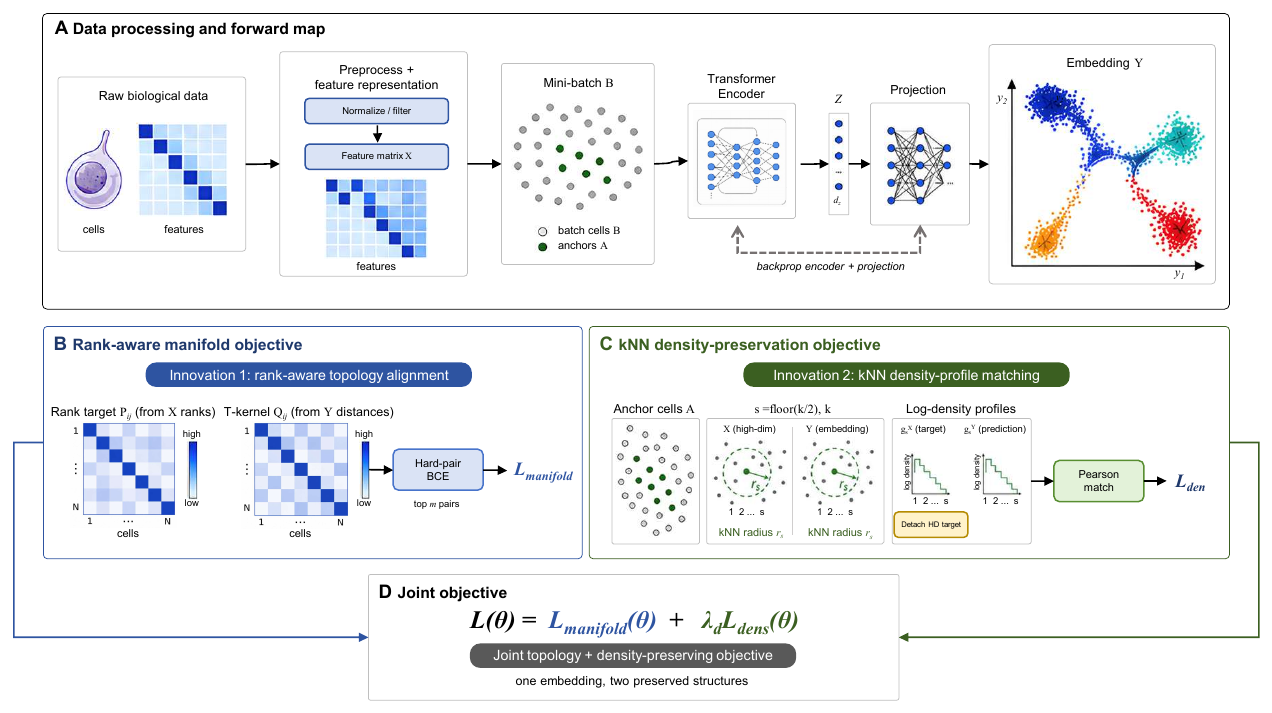}
\caption{DMT-Dens architecture and training objective. (A) A latent-token Transformer encoder and projection module map the input matrix $\mathbf{X}$ to the embedding $\mathbf{Y}$. For each observation, the encoder forms a fixed set of latent tokens, applies self-attention within the token set, and pools the result into a 40-dimensional representation for two-dimensional projection. (B) The manifold-preservation loss derives bidirectional rank affinities $P_{ij}$ from paired input-space views. Binary cross-entropy matches these targets to cross-view Student $t$-kernel affinities $Q_{ij}$ over selected hard pairs. (C) The density-preservation loss compares kNN log-density profiles $g_s^{X}$ and $g_s^{Y}$ across anchor observations at scales $s\in\{\lfloor k/2\rfloor,k\}$. Stop-gradient is applied to the input-space profile. (D) The final objective is $\mathcal{L}(\theta)=\mathcal{L}_{\mathrm{manifold}}(\theta)+\lambda_d\mathcal{L}_{\mathrm{dens}}(\theta)$; its gradients update the encoder and projection module.\label{fig:model-construction}}
\end{figure*}

\subsection{Latent-token Transformer encoder}
\label{sec:latent-transformer}

The encoder $e_\phi$ represents each observation with a fixed number of latent tokens. Self-attention operates within this token set and is not applied directly across input features or observations. The attention sequence length is independent of the input dimension $D$, and the embedding of an observation does not depend on other observations in its inference batch. We first normalize each input using
\begin{equation}
\overline{\mathbf{x}}_i=\operatorname{BN}_{\mathrm{in}}(\mathbf{x}_i).
\end{equation}
A learned linear map compresses the normalized feature vector into $M$ latent vectors of width $r$,
\begin{equation}
\mathbf{S}_i=\operatorname{reshape}\!\left(
\mathbf{W}_{c}\overline{\mathbf{x}}_i,\,M,\,r
\right)\in\mathbb{R}^{M\times r},
\end{equation}
where $\mathbf{W}_{c}\in\mathbb{R}^{Mr\times D}$. Each vector is independently expanded to the token width $d_t$ and assigned a learned latent-identity embedding,
\begin{equation}
\mathbf{T}^{(0)}_{i,m}
=\mathbf{S}_{i,m}\mathbf{E}_{m}+\mathbf{a}_{m},
\qquad m=1,\dots,M,
\end{equation}
where $\mathbf{E}_{m}\in\mathbb{R}^{r\times d_t}$ and $\mathbf{a}_{m}\in\mathbb{R}^{d_t}$. The resulting token matrix $\mathbf{T}^{(0)}_i\in\mathbb{R}^{M\times d_t}$ has the same size for datasets with different feature dimensions.

The tokens are processed by $L$ pre-normalized Transformer blocks \citep{vaswani2017attention}. For block $\ell$, the updates are
\begin{align}
\mathbf{U}^{(\ell)}_i
&=\mathbf{T}^{(\ell-1)}_i
+\operatorname{MSA}_{\ell}\!\left(
\operatorname{LN}_{\ell,1}(\mathbf{T}^{(\ell-1)}_i)
\right),\\
\mathbf{T}^{(\ell)}_i
&=\mathbf{U}^{(\ell)}_i
+\operatorname{FFN}_{\ell}\!\left(
\operatorname{LN}_{\ell,2}(\mathbf{U}^{(\ell)}_i)
\right),
\end{align}
where $\operatorname{MSA}$ denotes multi-head self-attention over the $M$ tokens of one observation, and $\operatorname{FFN}$ is a two-layer feed-forward network with GELU activation. The encoder uses neither labels nor cross-observation attention. After the final block, we average the layer-normalized tokens and map the pooled vector to 40 dimensions,
\begin{align}
\mathbf{h}_i
&=\frac{1}{M}\sum_{m=1}^{M}
\operatorname{LN}_{\mathrm{out}}\!\left(
\mathbf{T}^{(L)}_{i,m}
\right),\\
\mathbf{z}_i
&=\operatorname{BN}_{\mathrm{out}}\!\left(
\mathbf{W}_{o}\mathbf{h}_i+\mathbf{b}_{o}
\right).
\end{align}
The projection module produces $\mathbf{y}_i=p_\psi(\mathbf{z}_i)$. Our final configuration uses $M=32$, $r=16$, $d_t=224$, $L=2$, and four attention heads. The feed-forward hidden width is $4d_t$; attention and residual dropout are set to zero. Attention cost is determined by the fixed token count $M$ and does not scale quadratically with the original feature dimension $D$.

\subsection{Manifold-preservation loss}
\label{sec:manifold-loss}

Each mini-batch yields two paired views,
\begin{equation}
\mathcal{B}^{(a)}=\{\mathbf{x}^{(a)}_i\}_{i=1}^{B},
\qquad
\mathcal{B}^{(b)}=\{\mathbf{x}^{(b)}_i\}_{i=1}^{B},
\end{equation}
where $\mathbf{x}^{(a)}_i=\mathbf{x}_i$ is the original observation and $\mathbf{x}^{(b)}_i$ is a neighborhood-augmented version. To form the augmented view, we project the full preprocessed dataset onto $D_{\mathrm{pca}}$ principal components and build an approximate $K_{\mathrm{aug}}$-nearest-neighbor graph after excluding self-matches. These PCA coordinates are used only for neighbor search. For each sampled observation, we draw an index $j_i^{\mathrm{aug}}$ uniformly from its $K_{\mathrm{aug}}$ graph neighbors and interpolate in the original preprocessed feature space,
\begin{equation}
\mathbf{x}^{(b)}_i=\alpha_i\,\mathbf{x}_i+(1-\alpha_i)\,\mathbf{x}_{j_i^{\mathrm{aug}}},
\qquad
\alpha_i\sim\mathrm{Uniform}(\alpha_{\min},\alpha_{\max}).
\label{eq:augmentation}
\end{equation}
The vectors $\mathbf{x}^{(a)}_i$, $\mathbf{x}_{j_i^{\mathrm{aug}}}$, and $\mathbf{x}^{(b)}_i$ all lie in the original $D$-dimensional preprocessed feature space. PCA does not alter the vectors used in the distance calculations below. Each observation contributes one augmented vector, giving $B$ vectors in each view. For the nine-dataset benchmark, $K_{\mathrm{aug}}=200$, $D_{\mathrm{pca}}=64$, $\alpha_{\min}=0.05$, and $\alpha_{\max}=1$. Parameter settings that differ in the case studies are reported in the Supplementary Material. The shared mapping $f_\theta$ processes both views to produce paired two-dimensional embeddings,
\begin{equation}
\mathbf{y}^{(a)}_i=f_\theta(\mathbf{x}^{(a)}_i),
\qquad
\mathbf{y}^{(b)}_i=f_\theta(\mathbf{x}^{(b)}_i).
\end{equation}
Before computing embedding-space affinities, we standardize each coordinate within each view. For view $v\in\{a,b\}$ and coordinate $r\in\{1,2\}$, let $\widetilde{y}^{(v)}_{i,r}=(y^{(v)}_{i,r}-\mu^{(v)}_{r})/\sigma^{(v)}_{r}$, where $\mu^{(v)}_{r}$ and $\sigma^{(v)}_{r}$ are the mini-batch mean and standard deviation. These statistics are held constant during backpropagation. We write the standardized embedding as $\widetilde{\mathbf{y}}^{(v)}_i=(\widetilde{y}^{(v)}_{i,1},\widetilde{y}^{(v)}_{i,2})^\top$.

We define the raw cross-view squared Euclidean distance in the input space as
\begin{equation}
\Delta^{ab}_{ij}=\left\lVert\mathbf{x}^{(a)}_i-\mathbf{x}^{(b)}_j\right\rVert_2^2.
\end{equation}
The augmentation generally makes $\Delta^{ab}_{ii}$ nonzero. To treat matched observations as nearest across views, we floor off-diagonal squared distances at $\varepsilon_{\mathrm{dist}}>0$ and set the paired diagonal to zero:
\begin{equation}
d^{ab}_{ij}=
\begin{cases}
\max(\Delta^{ab}_{ij},\varepsilon_{\mathrm{dist}}), & i\neq j,\\
0, & i=j.
\end{cases}
\end{equation}
For each $\mathbf{x}^{(a)}_i$, let $\pi_i^{a\rightarrow b}(j)$ be the ordinal rank of $\mathbf{x}^{(b)}_j$ after sorting $\{d^{ab}_{ij}\}_{j=1}^{B}$ in ascending order. For a fixed $i$, the index $j$ ranges over the $B$ augmented vectors in the mini-batch. The $K_{\mathrm{aug}}$ candidates used to construct $\mathbf{x}^{(b)}_i$ are not part of this ranking. Ranks are zero-based, so $\pi_i^{a\rightarrow b}(j)\in\{0,\dots,B-1\}$. For tied off-diagonal distances, \texttt{torch.argsort} assigns distinct consecutive ranks in its returned order; average ranks are not used. Since $d^{ab}_{ii}=0$ and all off-diagonal entries are positive, the paired diagonal has rank $0$. The reverse rank $\pi_j^{b\rightarrow a}(i)$ is defined in the same way by sorting distances from $\mathbf{x}^{(b)}_j$ to the observations in the first view.

The directional rank affinities are defined as
\begin{equation}
P^{a\rightarrow b}_{ij}=\eta^{\,\pi_i^{a\rightarrow b}(j)},
\qquad
P^{b\rightarrow a}_{ij}=\eta^{\,\pi_j^{b\rightarrow a}(i)},
\end{equation}
where $\eta\in(0,1)$ controls affinity decay with rank. The paired diagonal has rank $0$ in both directions, giving $P^{a\rightarrow b}_{ii}=P^{b\rightarrow a}_{ii}=1$. We combine the two directional affinities using their geometric mean,
\begin{equation}
P_{ij}=\max\!\left(
\sqrt{P^{a\rightarrow b}_{ij}P^{b\rightarrow a}_{ij}},
\varepsilon_{\mathrm{aff}}
\right),
\end{equation}
where $\varepsilon_{\mathrm{aff}}>0$ is a numerical floor. This construction gives $P_{ii}=1$ and encodes the bidirectional neighborhood ranks of cross-view pairs. The affinity matrix $P$ is held fixed during optimization.

Let
\begin{equation}
\delta^{ab}_{ij}=\left\lVert
\widetilde{\mathbf{y}}^{(a)}_i-
\widetilde{\mathbf{y}}^{(b)}_j
\right\rVert_2
\end{equation}
denote the distance between the standardized embeddings of the two views. The corresponding Student $t$-kernel value is
\begin{equation}
\widetilde{Q}_{ij}=\left(
1+\frac{(\delta^{ab}_{ij})^2}{\nu}
\right)^{-(\nu+1)/2},
\end{equation}
where $\nu>0$ controls the tail weight of the kernel. The embedding-space affinity is obtained by row normalization,
\begin{equation}
Q_{ij}=
\frac{\widetilde{Q}_{ij}}
{\sum_{\ell\neq i}\widetilde{Q}_{i\ell}}.
\end{equation}

Pairwise binary cross-entropy matches the input-space targets to the embedding-space affinities,
\begin{equation}
\ell_{ij}=-\left[
P_{ij}\log(Q_{ij}+\varepsilon_{\mathrm{BCE}})
+(1-P_{ij})\log(1-Q_{ij}+\varepsilon_{\mathrm{BCE}})
\right],
\end{equation}
where $\varepsilon_{\mathrm{BCE}}>0$ stabilizes the logarithms. The pointwise loss measures the mismatch between an input-space rank affinity and its embedding-space counterpart.

Hard-pair selection retains the cross-view pairs with the largest pointwise losses. For each $i$, let $\tau_i$ be the $m$-th largest value in $\{\ell_{ij}\}_{j=1}^{B}$, where $m=\min(m_0,B)$, and define $\mathcal{M}_i=\{\,j:\ell_{ij}\ge\tau_i\,\}$. We use $|\cdot|$ for set cardinality, so $|\mathcal{M}_i|$ is the number of indices retained for observation $i$. The threshold gives $|\mathcal{M}_i|\ge m$, with equality when no losses are tied at $\tau_i$. The full set of retained pairs is
\begin{equation}
\mathcal{M}=\{(i,j):j\in\mathcal{M}_i\}.
\end{equation}
The manifold-preservation loss is then defined as
\begin{equation}
\mathcal{L}_{\mathrm{manifold}}
=\frac{1}{|\mathcal{M}|}\sum_{(i,j)\in\mathcal{M}}\ell_{ij}.
\end{equation}
The mini-batch loss is consequently concentrated on cross-view relationships with the largest current mismatch between $P_{ij}$ and $Q_{ij}$.

The manifold-preservation loss integrates bidirectional rank targets, Student $t$-kernel similarities, and hard-pair aggregation. It is optimized jointly with the density-preservation loss in Eq.~\eqref{eq:total-loss-overview}.

\subsection{Density-preservation loss}
\label{sec:density-loss}
We estimate local density using $k$-nearest-neighbor (kNN) radii. For a reference index set $\mathcal{R}$ and a query $a\in\mathcal{R}$, define the input- and embedding-space radii as
\begin{equation}
\begin{aligned}
r_s^{X}(a,\mathcal{R})
&=\operatorname{OS}_{s}\!\left(
\left\{\lVert\mathbf{x}_a-\mathbf{x}_j\rVert_2:
j\in\mathcal{R}\setminus\{a\}\right\}\right),\\
r_s^{Y}(a,\mathcal{R})
&=\operatorname{OS}_{s}\!\left(
\left\{\lVert\mathbf{y}_a-\mathbf{y}_j\rVert_2:
j\in\mathcal{R}\setminus\{a\}\right\}\right),
\end{aligned}
\end{equation}
where $\operatorname{OS}_{s}$ denotes the $s$-th smallest value. The query is excluded from both reference neighborhoods. In a space of ambient dimension $d$, let $n_{\mathcal{R}}=|\mathcal{R}|-1$ be the number of eligible reference observations and $V_d=\pi^{d/2}/\Gamma(d/2+1)$ the volume of the unit $d$-ball. The classical estimator is $\hat{f}_k(\mathbf{x}_a)=k/[n_{\mathcal{R}}V_d r_k(\mathbf{x}_a,\mathcal{R})^d]$; in the input space, $r_k(\mathbf{x}_a,\mathcal{R})=r_k^{X}(a,\mathcal{R})$. Supplementary Section~\ref{sec:supp-density-derivation} gives the derivation. Our density loss uses the Pearson correlation between input- and embedding-space log-density vectors. Under this correlation, the terms for sample size, neighbor count, and ambient dimension contribute only an additive constant and a positive scale factor. We use the resulting dimension-free quantity
\begin{equation}
\log\hat{f}_k(\mathbf{x}_a)\,\propto\,-\log r_k^{X}(a,\mathcal{R}).
\label{eq:dimfree}
\end{equation}
In the implementation, the zero self-distance occupies the first position in each query-to-reference distance row, so the radius is the $(s{+}1)$-th smallest entry.

\subsection{Training objective and optimization}

The dimension-free log-density in Eq.~\eqref{eq:dimfree} increases from sparse to dense regions. We compare the input- and embedding-space profiles using Pearson correlation. Within a mini-batch of $B$ observations, the correlation is evaluated on a random anchor subset $\mathcal{A}\subseteq\{1,\dots,B\}$ of size $A=\min(A_0,B)$. We use two neighborhood scales, $\mathcal{S}=\{\lfloor k/2\rfloor,\,k\}$.

Let $\mathcal{R}_B=\{1,\dots,B\}$ be the common reference index set for the original observations and their current embeddings. For each anchor $a\in\mathcal{A}$ and scale $s\in\mathcal{S}$, define
\begin{align}
g^{X}_{s,a}&=\operatorname{sg}\!\left[
-\log\!\left(r_s^{X}(a,\mathcal{R}_B)+\varepsilon_{\log}\right)
\right],\\
g^{Y}_{s,a}&=-\log\!\left(r_s^{Y}(a,\mathcal{R}_B)+\varepsilon_{\log}\right).
\end{align}
Here, $\varepsilon_{\log}>0$ stabilizes the logarithm. The stop-gradient operator satisfies $\operatorname{sg}[u]=u$ in the forward pass and $\partial\operatorname{sg}[u]/\partial u=0$. Let $\mathbf{g}^{X}_{s}=(g^{X}_{s,a})_{a\in\mathcal{A}}$ and $\mathbf{g}^{Y}_{s}=(g^{Y}_{s,a})_{a\in\mathcal{A}}$ be the anchor-indexed log-density vectors at scale $s$, and let $\rho(\mathbf{u},\mathbf{v})$ denote their Pearson correlation. The density-preservation loss averages one minus this correlation across scales,
\begin{equation}
\mathcal{L}_{\mathrm{dens}}
=\frac{1}{|\mathcal{S}|}\sum_{s\in\mathcal{S}}
 \Bigl[\,1-\rho\bigl(\mathbf{g}^{Y}_{s},\,\mathbf{g}^{X}_{s}\bigr)\Bigr].
\end{equation}
We compute the correlation by centering each vector and normalizing by its $\ell_2$ norm: $\rho(\mathbf{u},\mathbf{v})=\langle\mathbf{u}-\bar{u},\,\mathbf{v}-\bar{v}\rangle\big/\bigl(\lVert\mathbf{u}-\bar{u}\rVert\,\lVert\mathbf{v}-\bar{v}\rVert+\varepsilon_\rho\bigr)$. A single-scale version retains only $s=k$; the ablation study evaluates the contribution of using both scales in $\mathcal{S}$.

We jointly optimize the manifold- and density-preservation losses using Eq.~\eqref{eq:total-loss-overview}, with density weight $\lambda_d\ge 0$.

\subsection{Model implementation}
\label{sec:model-implementation}

Unless noted otherwise, we train each model for 1000 epochs with AdamW, an initial learning rate of $1\times10^{-3}$, cosine learning-rate annealing, mixed precision, and a batch size of 4096. The benchmark configuration uses $A_0=512$, $\lambda_d=1.8\times10^{-3}$, and $k=12$. Complete implementation settings and computational measurements are reported in the Supplementary Material.

\subsection{Datasets}
\label{sec:datasets}

The main benchmark contains nine datasets: a synthetic branching dataset, four scRNA-seq datasets, two image datasets, a text dataset, and a sensor dataset. Supplementary Table~\ref{tab:datasets} summarizes these datasets together with the \emph{C.\ elegans} (CELEGAN) developmental case study. The biological data comprise HCL \citep{han2020hcl}, MCA \citep{han2018mca}, GAST10K \citep{zhang2019gastric}, EPI, and CELEGAN \citep{packer2019celegan}; they cover cell-type atlases, lesion-associated epithelial states, and developmental trajectories. We reserve CELEGAN for the developmental case study because it includes both cell-type and embryonic-time annotations. A separately generated synthetic \texttt{dyngen} trajectory appears only in Supplementary Section~\ref{sec:supp-dyngen}.

\subsection{Baselines and evaluation metrics}
\label{sec:evaluation-setup}

We compare DMT-Dens with six dimensionality-reduction methods: t-SNE \citep{maaten2008visualizing}, UMAP \citep{mcinnes2018umap,becht2019umap}, PaCMAP \citep{wang2021pacmap}, PHATE \citep{moon2019phate}, den-SNE, and densMAP \citep{narayan2021assessing}. Baseline settings and hyperparameter search grids are reported in the Supplementary Material.

The main evaluation considers density preservation and label separability. Following the convention used for den-SNE and densMAP \citep{narayan2021assessing}, density correlation is the Spearman correlation between local radii in the high-dimensional space and the embedding. Each radius is the mean distance to the $k$ nearest neighbors. The evaluation is computed on the full dataset. The training loss uses $k$-th-neighbor log-radii and Pearson correlation evaluated on mini-batch anchors at two scales. Label separability is measured by the stratified five-fold cross-validated accuracy of a linear support-vector classifier (SVC) fitted to the two-dimensional embedding. The Supplementary Material reports additional measures of local fidelity, global distance, runtime, and biological trajectory structure.

\subsection{Ablation design}
\label{sec:ablation-design}

We conducted single-factor ablations on EPI, HCL, and MNIST. Relative to the full model, four variants use distance affinity, unidirectional matching, all-pair averaging, or only the density scale $s=k$. A fifth variant removes density regularization. All other settings, including the architecture, augmentation, optimizer, batch size, 1000-epoch schedule, and dataset-specific hyperparameters, are held fixed. Each variant uses the same three seeds. We report density correlation, kNN preservation, and SVC accuracy.

%% file: sections/results.tex
\section{Results}
\subsection{Quantitative results}
Table~\ref{tab:density-local-svc-main} compares the density correlation and SVC accuracy of DMT-Dens and six baseline methods across nine benchmark datasets. Results are reported over five seeds, and den-SNE is marked OOT on EMNIST because its runtime was 24\,h+. Local density correlation and additional embedding-quality metrics are provided in the Supplementary Material.

\begin{table*}[t]
\caption{Main comparison on density preservation and label separability. Higher is better for both metrics, and the best (highest) value in each row is shown in \textbf{bold}. Entries report mean$\pm$standard deviation over available seeds. The final row of each panel reports the unweighted mean over all nine datasets; this mean is omitted for den-SNE because its EMNIST run did not complete. OOT (out of time) denotes a runtime of 24\,h+; this applies to den-SNE on the full EMNIST (\texttt{byclass}, ${\approx}698$K points) benchmark.\label{tab:density-local-svc-main}}
\centering
\scriptsize
\setlength{\tabcolsep}{3pt}
\renewcommand{\arraystretch}{1.04}
\setlength{\tabcolsep}{2pt}
\textbf{A. Density correlation}\par\vspace{2pt}
\begin{tabular*}{\textwidth}{@{\extracolsep{\fill}}llccccccc@{\extracolsep{\fill}}}
\toprule
Type & Dataset & t-SNE & UMAP & den-SNE & densMAP & PaCMAP & PHATE & DMT-Dens \\
\midrule
\multirow{1}{*}{\textit{Synthetic}}
 & ArtificialTree & $0.812\pm0.010$ & $0.626\pm0.034$ & $0.575\pm0.031$ & $0.746\pm0.031$ & $0.653\pm0.028$ & $-0.188\pm0.098$ & \best{$0.954\pm0.004$} \\
\midrule
\multirow{4}{*}{\textit{Biological}}
 & HCL & $0.110\pm0.020$ & $0.036\pm0.031$ & $0.718\pm0.017$ & $0.531\pm0.015$ & $0.083\pm0.031$ & $0.382\pm0.009$ & \best{$0.870\pm0.003$} \\
 & GAST10K & $0.061\pm0.038$ & $-0.156\pm0.017$ & $0.786\pm0.011$ & $0.798\pm0.015$ & $-0.530\pm0.019$ & $0.169\pm0.020$ & \best{$0.872\pm0.008$} \\
 & EPI & $0.222\pm0.035$ & $0.086\pm0.029$ & $0.705\pm0.011$ & $0.712\pm0.012$ & $0.124\pm0.025$ & $0.243\pm0.016$ & \best{$0.792\pm0.014$} \\
 & MCA & $0.416\pm0.025$ & $0.112\pm0.012$ & $0.327\pm0.015$ & $0.466\pm0.019$ & $-0.001\pm0.032$ & $-0.028\pm0.031$ & \best{$0.644\pm0.068$} \\
\midrule
\multirow{4}{*}{\textit{Non-biological}}
 & MNIST & $0.296\pm0.048$ & $-0.081\pm0.020$ & \best{$0.768\pm0.014$} & $0.756\pm0.020$ & $-0.020\pm0.040$ & $0.152\pm0.022$ & $0.700\pm0.028$ \\
 & EMNIST & $0.241\pm0.025$ & $0.025\pm0.025$ & OOT & $0.600\pm0.022$ & $0.094\pm0.018$ & $0.166\pm0.043$ & \best{$0.678\pm0.015$} \\
 & NG20 & $0.386\pm0.029$ & $0.281\pm0.021$ & $0.758\pm0.007$ & \best{$0.797\pm0.010$} & $0.190\pm0.037$ & $0.609\pm0.015$ & $0.674\pm0.017$ \\
 & ACT & $0.355\pm0.051$ & $0.147\pm0.037$ & \best{$0.902\pm0.007$} & $0.840\pm0.011$ & $0.135\pm0.043$ & $0.394\pm0.093$ & $0.832\pm0.007$ \\
\midrule
\multicolumn{2}{l}{Mean} & $0.322$ & $0.120$ & -- & $0.694$ & $0.081$ & $0.211$ & \best{$0.780$} \\
\botrule
\end{tabular*}

\vspace{4pt}
\textbf{B. SVC accuracy}\par\vspace{2pt}
\begin{tabular*}{\textwidth}{@{\extracolsep{\fill}}llccccccc@{\extracolsep{\fill}}}
\toprule
Type & Dataset & t-SNE & UMAP & den-SNE & densMAP & PaCMAP & PHATE & DMT-Dens \\
\midrule
\multirow{1}{*}{\textit{Synthetic}}
 & ArtificialTree & \best{$0.848\pm0.019$} & $0.548\pm0.022$ & $0.589\pm0.035$ & $0.610\pm0.020$ & $0.672\pm0.026$ & $0.227\pm0.064$ & $0.796\pm0.022$ \\
\midrule
\multirow{4}{*}{\textit{Biological}}
 & HCL & $0.601\pm0.020$ & $0.304\pm0.019$ & $0.752\pm0.032$ & $0.282\pm0.004$ & $0.755\pm0.014$ & $0.337\pm0.029$ & \best{$0.781\pm0.015$} \\
 & GAST10K & $0.654\pm0.014$ & $0.597\pm0.026$ & $0.770\pm0.051$ & $0.593\pm0.013$ & $0.772\pm0.024$ & $0.736\pm0.016$ & \best{$0.792\pm0.031$} \\
 & EPI & $0.839\pm0.012$ & $0.848\pm0.010$ & $0.831\pm0.026$ & $0.843\pm0.010$ & $0.858\pm0.012$ & $0.788\pm0.004$ & \best{$0.884\pm0.022$} \\
 & MCA & $0.435\pm0.018$ & $0.326\pm0.025$ & \best{$0.807\pm0.023$} & $0.290\pm0.014$ & $0.603\pm0.031$ & $0.713\pm0.016$ & $0.774\pm0.059$ \\
\midrule
\multirow{4}{*}{\textit{Non-biological}}
 & MNIST & $0.931\pm0.006$ & \best{$0.962\pm0.003$} & $0.922\pm0.007$ & $0.945\pm0.004$ & $0.959\pm0.002$ & $0.706\pm0.032$ & $0.950\pm0.007$ \\
 & EMNIST & $0.601\pm0.020$ & \best{$0.655\pm0.008$} & OOT & $0.646\pm0.012$ & $0.626\pm0.012$ & $0.468\pm0.005$ & $0.639\pm0.008$ \\
 & NG20 & $0.198\pm0.033$ & $0.219\pm0.008$ & $0.294\pm0.021$ & $0.264\pm0.017$ & $0.245\pm0.021$ & $0.239\pm0.019$ & \best{$0.297\pm0.034$} \\
 & ACT & $0.859\pm0.007$ & $0.823\pm0.019$ & $0.831\pm0.016$ & $0.811\pm0.032$ & $0.842\pm0.015$ & $0.783\pm0.021$ & \best{$0.876\pm0.003$} \\
\midrule
\multicolumn{2}{l}{Mean} & $0.663$ & $0.587$ & -- & $0.587$ & $0.704$ & $0.555$ & \best{$0.754$} \\
\botrule
\end{tabular*}
\end{table*}

Based on the reported means, DMT-Dens had the highest density correlation on ArtificialTree, HCL, GAST10K, EPI, MCA, and EMNIST. den-SNE had the highest value on MNIST and ACT, and densMAP had the highest value on NG20. DMT-Dens ranked first or second in mean SVC accuracy on seven datasets and third on MNIST and EMNIST.

Figure~\ref{fig:motivation}c reports the mean density-correlation and SVC-accuracy ranks over the eight datasets with complete results for all methods. EMNIST was excluded because the den-SNE runtime was 24\,h+. DMT-Dens had the lowest mean rank for both metrics and lay on the Pareto frontier when all seven methods were considered. Among the baselines, densMAP and den-SNE had the two lowest mean density-correlation ranks, while PaCMAP and den-SNE had the two lowest mean SVC-accuracy ranks.

\subsection{Biological embedding visualizations}
\begin{figure*}[p]
\centering
\includegraphics[width=0.85\textwidth,height=0.49\textheight,keepaspectratio]{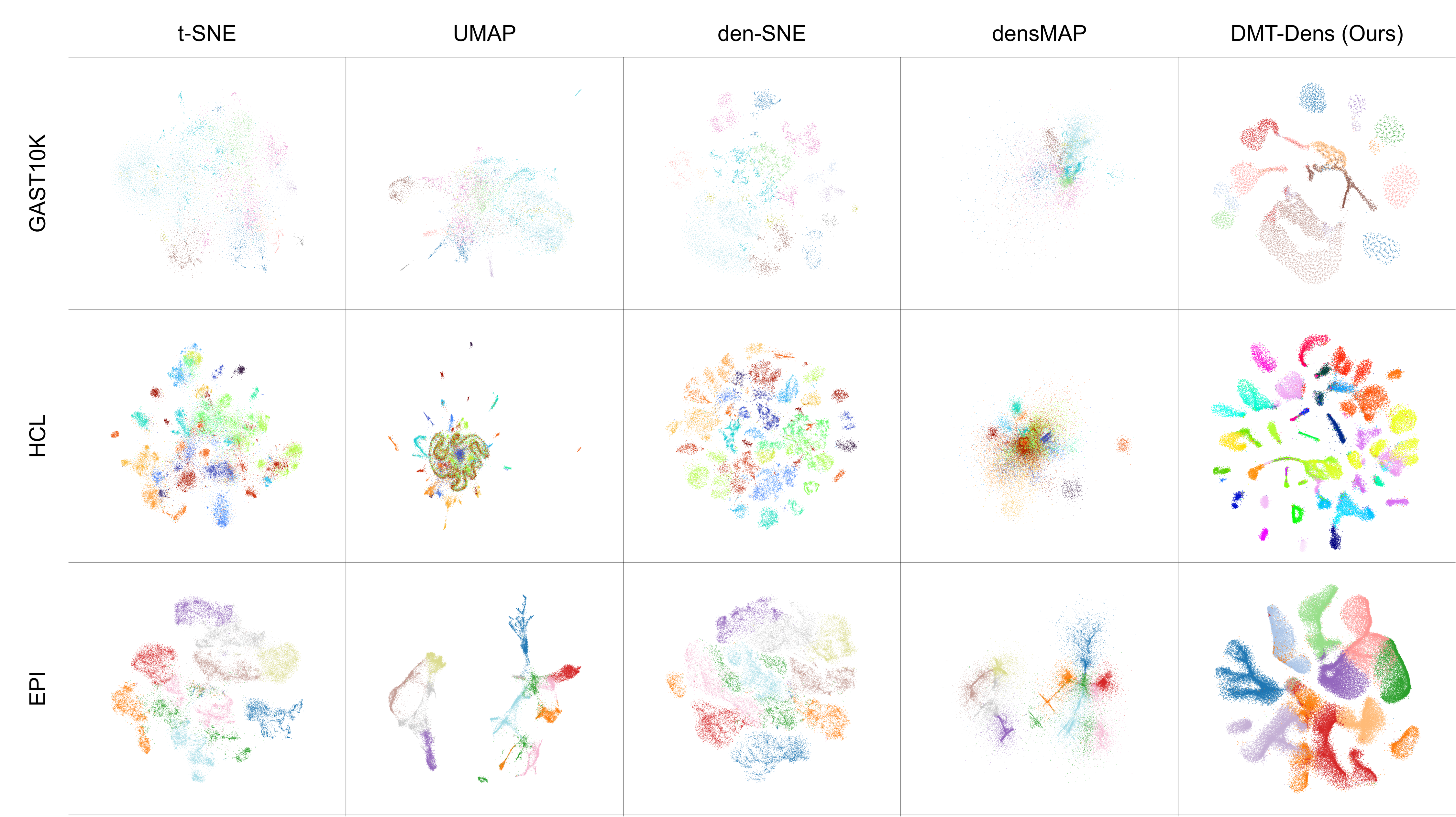}
\caption{Two-dimensional embeddings of three single-cell datasets. Rows correspond to GAST10K, HCL, and EPI, and columns correspond to t-SNE, UMAP, den-SNE, densMAP, and DMT-Dens. Points are colored by cell type, and axis limits are matched across methods within each dataset. DMT-Dens is shown in the rightmost column. Supplementary Figure~\ref{fig:comparison-full} includes MCA, PHATE, and PaCMAP.\label{fig:comparison}}
\vspace{6pt}
\includegraphics[width=0.85\textwidth,height=0.40\textheight,keepaspectratio]{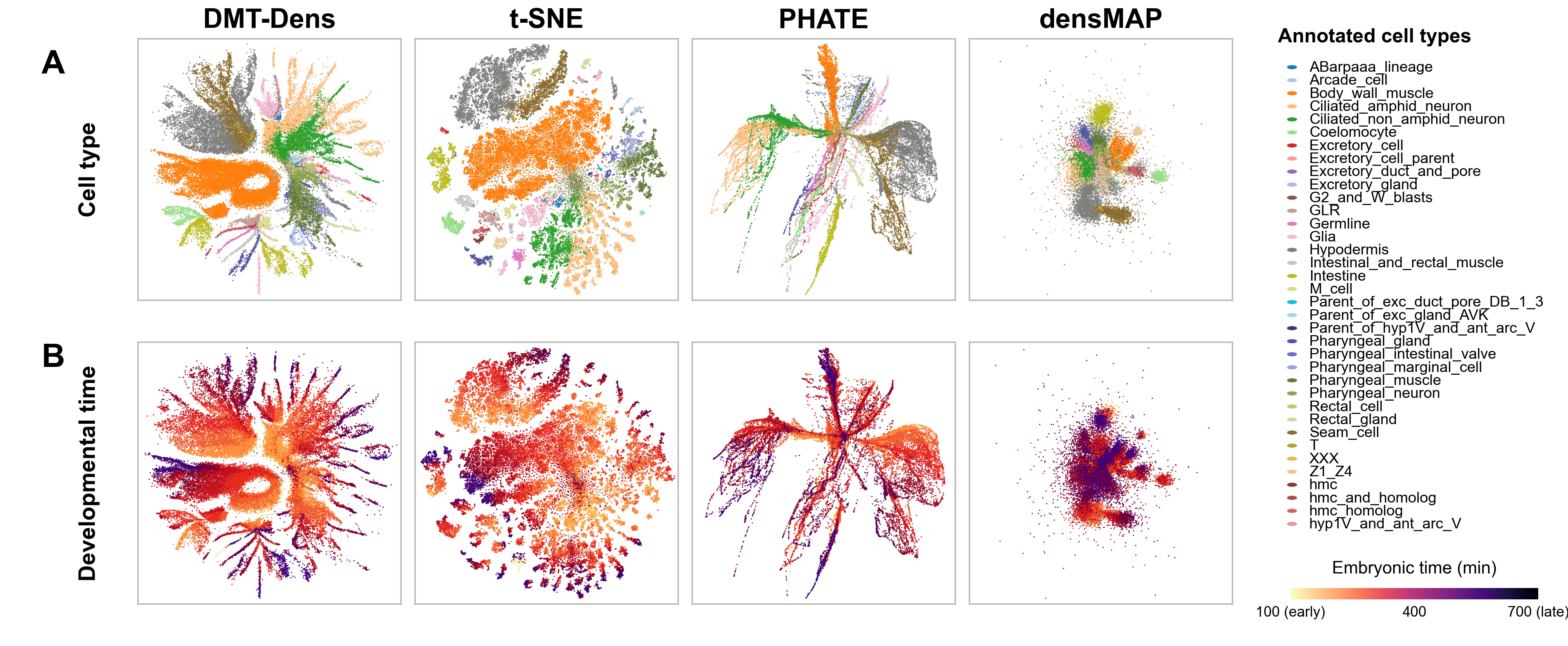}
\caption{Two-dimensional embeddings of the \emph{C.\ elegans} embryonic time-course data produced by DMT-Dens, t-SNE, PHATE, and densMAP. \textbf{(A)}~Embeddings colored by annotated cell type. \textbf{(B)}~The same embeddings colored by observed developmental time. Quantitative developmental-time metrics are reported in Supplementary Table~\ref{tab:supp-celegan-traj}.\label{fig:case-study}}
\end{figure*}

Figure~\ref{fig:comparison} displays the two-dimensional embeddings produced by DMT-Dens, t-SNE, UMAP, den-SNE, and densMAP for GAST10K, HCL, and EPI. Points are colored by the cell-type annotations used for evaluation. Within the DMT-Dens panels, local point density differs across annotated populations, with both compact and diffuse regions visible in GAST10K, HCL, and EPI. The full seven-method comparison for all datasets is provided in Supplementary Figure~\ref{fig:comparison-full}.

\subsection{Ablation results}
Table~\ref{tab:ablation-main} reports the ablations described in Section~\ref{sec:ablation-design}. Across the three datasets, all-pair averaging produced the highest mean density correlation, together with the lowest mean kNN preservation and SVC accuracy. Removing density regularization produced the opposite pattern: density correlation was the lowest among the six settings, whereas kNN preservation and SVC accuracy were higher than those of the full model. This variant attained the highest kNN preservation on MNIST and the highest SVC accuracy on HCL and MNIST. Replacing rank affinity with distance affinity lowered all three metrics relative to the full model. The unidirectional variant increased kNN preservation on every dataset and attained the highest values on EPI and HCL, while its density correlation remained below that of the full model. The single-scale-density variant also had lower density correlation than the full model and its SVC accuracy increased on EPI and decreased slightly on HCL and MNIST.

Using the mean of density correlation and SVC accuracy, the full model obtained the highest value among the six settings on EPI, HCL, and MNIST. No ablation variant exceeded the full model on all three reported metrics for any dataset.

\begingroup
\setlength{\intextsep}{8pt}
\begin{table*}[!t]
\caption{Ablation results on EPI, HCL, and MNIST. Each row changes one component of the full model at a time, as defined in Section~\ref{sec:ablation-design}. Dens., kNN, and SVC denote density correlation, kNN preservation, and SVC accuracy, respectively. Entries report mean$\pm$sample standard deviation over three matched seeds (42--44). Higher is better, and the highest value in each dataset--metric column is shown in \textbf{bold}.\label{tab:ablation-main}}
\centering
\scriptsize
\setlength{\tabcolsep}{2.2pt}
\renewcommand{\arraystretch}{1.05}
\makeatletter\def\cline#1{\@cline#1\@nil}\makeatother
\resizebox{\textwidth}{!}{%
\begin{tabular}{@{}lccccccccc@{}}
\toprule
Setting & \multicolumn{3}{c}{EPI} & \multicolumn{3}{c}{HCL} & \multicolumn{3}{c}{MNIST} \\
\cline{2-4}\cline{5-7}\cline{8-10}
\noalign{\vskip2pt}
 & Dens. & kNN & SVC & Dens. & kNN & SVC & Dens. & kNN & SVC \\
\midrule
\multicolumn{10}{@{}c@{}}{{\setlength{\fboxsep}{0pt}\colorbox{tablegroupgray}{\makebox[\textwidth][c]{\rule[-0.45ex]{0pt}{2.4ex}\textit{Remove or replace one component of Full}}}}} \\
Full & $0.817\pm0.001$ & $0.155\pm0.002$ & $0.885\pm0.019$ & $0.863\pm0.003$ & $0.092\pm0.001$ & $0.790\pm0.030$ & $0.697\pm0.003$ & $0.310\pm0.002$ & $0.950\pm0.001$ \\
Distance affinity & $0.674\pm0.002$ & $0.108\pm0.000$ & $0.736\pm0.004$ & $0.818\pm0.004$ & $0.062\pm0.003$ & $0.485\pm0.010$ & $0.607\pm0.043$ & $0.250\pm0.003$ & $0.745\pm0.012$ \\
Unidirectional & $0.765\pm0.017$ & \best{$0.166\pm0.002$} & $0.870\pm0.048$ & $0.844\pm0.005$ & \best{$0.101\pm0.001$} & $0.801\pm0.002$ & $0.673\pm0.006$ & $0.313\pm0.001$ & $0.952\pm0.003$ \\
All-pair & \best{$0.905\pm0.002$} & $0.035\pm0.005$ & $0.447\pm0.011$ & \best{$0.960\pm0.002$} & $0.028\pm0.001$ & $0.380\pm0.026$ & \best{$0.896\pm0.011$} & $0.117\pm0.020$ & $0.691\pm0.067$ \\
Single-scale density & $0.791\pm0.011$ & $0.153\pm0.003$ & \best{$0.906\pm0.008$} & $0.860\pm0.004$ & $0.092\pm0.001$ & $0.788\pm0.006$ & $0.683\pm0.010$ & $0.309\pm0.001$ & $0.943\pm0.004$ \\
No density & $0.211\pm0.007$ & $0.162\pm0.001$ & $0.904\pm0.008$ & $0.068\pm0.034$ & $0.097\pm0.002$ & \best{$0.842\pm0.003$} & $-0.054\pm0.041$ & \best{$0.340\pm0.001$} & \best{$0.955\pm0.001$} \\
\botrule
\end{tabular}
}
\end{table*}
\endgroup

\subsection{Biological case study}
\begin{figure}[!t]
\centering
\includegraphics[width=\linewidth]{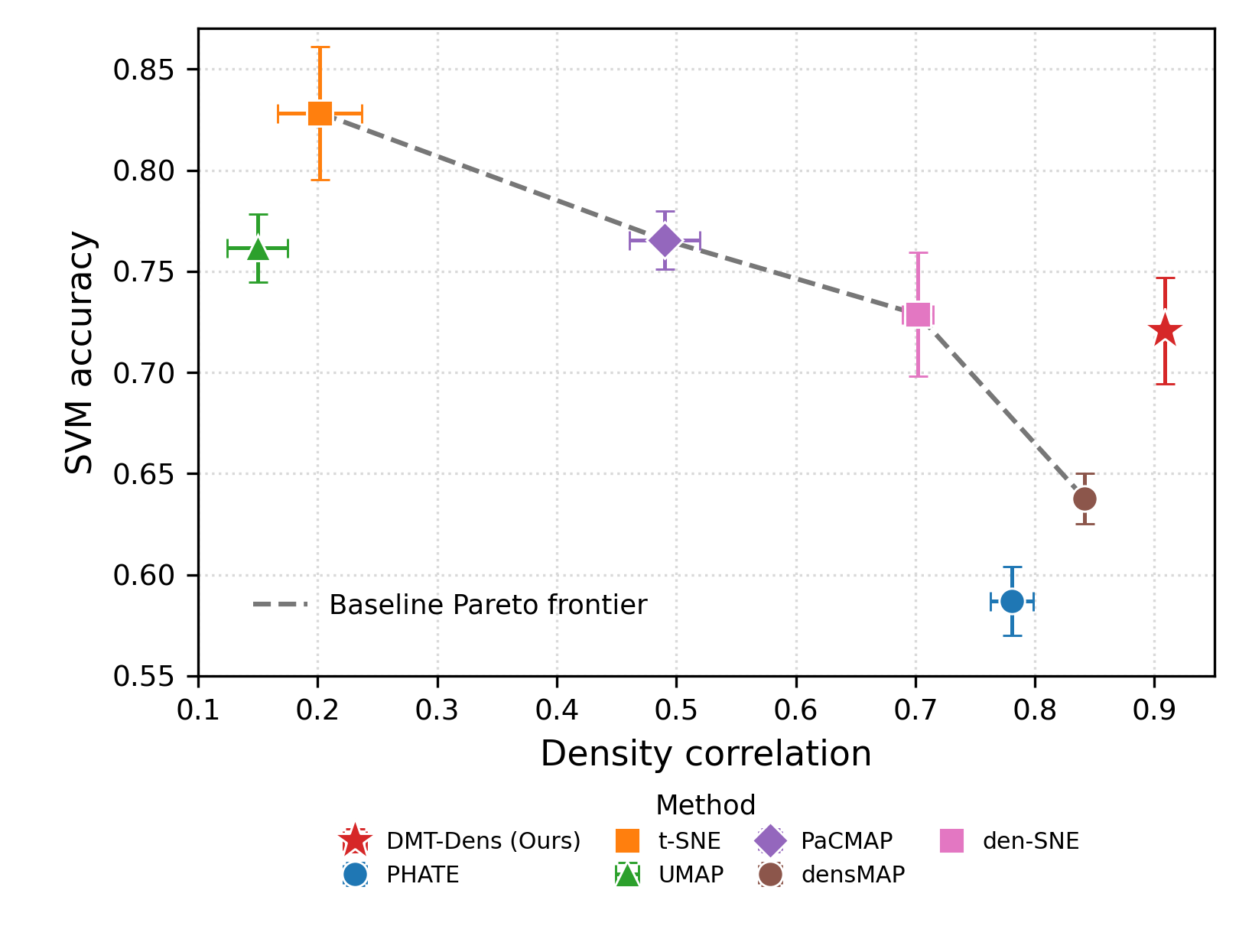}
\caption{Density correlation versus SVC accuracy on the \textit{C.\ elegans} case study (mean $\pm$ s.d.\ over five seeds). The dashed line marks the baseline Pareto frontier. DMT-Dens had a mean density correlation of $0.909$ and a mean SVC accuracy of $0.721$; the highest baseline values for these metrics were $0.841$ for densMAP and $0.828$ for t-SNE, respectively.\label{fig:case-study-metrics}}
\end{figure}

We evaluated DMT-Dens on the \emph{C.\ elegans} embryonic time course (CELEGAN) \citep{packer2019celegan}. Each cell has a cell-type annotation and an embryonic-time annotation recorded as a discrete interval in minutes after fertilization. Figure~\ref{fig:case-study} displays the embeddings colored separately by these two annotations. Figure~\ref{fig:case-study-metrics} reports density correlation and SVC accuracy. Supplementary Section~\ref{sec:supp-celegan-traj} and Supplementary Table~\ref{tab:supp-celegan-traj} report pseudotime correlation, ordering accuracy, time continuity, DEMaP, and the reachable fraction of cells. Pseudotime correlation and ordering accuracy are computed on the reachable subset when an embedding graph is fragmented, and the corresponding reachable fractions are reported in the same table.

\subsection{Supplementary results}
Supplementary Table~\ref{tab:supplementary-metrics} reports additional embedding metrics, and Supplementary Figures~\ref{fig:comparison-full} and~\ref{fig:text-comparison-full} show the full seven-method embedding comparisons. Supplementary Tables~\ref{tab:supp-param-anchor}--\ref{tab:supp-param-lambda} report the analyses of the density weight $\lambda_d$, neighborhood scale $k$, and number of density anchors. Runtime and peak GPU memory for DMT-Dens are reported in Supplementary Tables~\ref{tab:supp-runtime} and~\ref{tab:supp-memory-all}.

Supplementary Section~\ref{sec:supp-dyngen} reports results for all methods on synthetic single-cell data generated with dyngen \citep{cannoodt2021dyngen}. The simulation provides the trajectory topology, branch assignments, and simulation time.

%% file: sections/discussion.tex
\section{Discussion}
DMT-Dens integrates rank-affinity manifold alignment, hard-pair optimization, and an explicit density-consistency regularizer. Across the benchmark datasets, the resulting embeddings showed high density correlation, particularly on the biological datasets, while remaining competitive in SVC accuracy. The ablation results distinguish the roles of the objective components. Removing density regularization markedly reduced density correlation, whereas all-pair averaging increased density correlation but lowered kNN preservation and SVC accuracy. Unidirectional matching and single-scale density estimation produced dataset-dependent changes. No ablation improved all three metrics within any dataset, indicating that density fidelity, neighborhood preservation, and label separability are related but distinct objectives.

Nonlinear projections can alter local point density and the apparent extent of cell populations \citep{narayan2021assessing,chari2023specious}. The density-consistency term penalizes disagreement between density estimates in the processed input and the embedding, encouraging compact and diffuse regions in two dimensions to follow the corresponding input-space variation. For rare transition cells \citep{zhou2021dissecting} and continuous differentiation processes \citep{setty2019palantir}, this information complements neighborhood structure and annotation-based views of the embedding.

The meaning of the retained density requires care. DMT-Dens operates on the processed input data, whose density reflects experimental sampling and preprocessing. Tissue dissociation can also induce transcriptional changes \citep{vandenbrink2017single}. The area occupied by a population should therefore be interpreted as observed sample density, not as a calibrated estimate of biological abundance. This distinction is especially important for rare and transitional states, which are sensitive to sampling variation. Continuous cell-state density estimators such as Mellon \citep{otto2024mellon} offer a basis for evaluating alternative density definitions.

The current formulation retains two user-specified choices: the neighborhood scale and the density-loss weight. Their effects varied across datasets in the sensitivity analyses, motivating further study of adaptive neighborhood selection and data-dependent loss calibration. The dyngen evaluation also shows that density fidelity and trajectory recovery need not align: DMT-Dens led the density metrics, while t-SNE and den-SNE attained the highest Branch SVC and topology fidelity, respectively. Density-aware projections should therefore be evaluated alongside topology-specific criteria when trajectory structure is central to the analysis. The parametric mapping supports projection of new samples, and the density regularizer could be evaluated in other parametric projection models \citep{xu2023dmt}. Further work should also test its interaction with trajectory-inference \citep{saelens2019comparison,wolf2019paga} and RNA-velocity-based fate-mapping workflows \citep{bergen2020velocity,lange2022cellrank}.

DMT-Dens provides a parametric approach for retaining relative density patterns while balancing neighborhood structure and label separability. Its embeddings remain representations of sampled and preprocessed data, and their biological interpretation should explicitly account for technical and sampling effects. Evaluation on larger datasets and in downstream tasks is needed to establish when improved density fidelity changes biological analyses.

%% file: sections/backmatter.tex
\section*{Conflicts of interest}
The authors declare no competing interests.

\section*{Funding}
Funding information will be added in the final version.

\section*{Data availability}
All datasets used in this work are publicly available. Source code and reproducible scripts are available at \url{https://github.com/Ruizhe-wang/DMT-Dens}.

\section*{Author contributions statement}
Author contributions will be completed after the author list is finalized.

\section*{Acknowledgments}
Acknowledgments will be added in the final version.

%% file: sections/supplementary.tex
\section{Datasets}
\label{sec:supp-datasets}
Table~\ref{tab:datasets} summarizes the nine benchmark datasets and the \emph{C.\ elegans} (CELEGAN) case-study dataset described in the main-text Methods. The separately generated \texttt{dyngen} simulation is described in Section~\ref{sec:supp-dyngen}. The table reports the modality, sample size $n$, feature dimension $D$, number of class labels, and evaluation annotation for each listed dataset. Labels are excluded from representation training but are used in the combined hyperparameter-selection criterion and in downstream evaluation and visualization. The $K_{\mathrm{aug}}$-nearest-neighbor graph used for data augmentation is computed once in a 64-dimensional principal-component space. We use $K_{\mathrm{aug}}=200$ for the main nine-dataset benchmark, $K_{\mathrm{aug}}=50$ for the CELEGAN case study, and $K_{\mathrm{aug}}=500$ for the dyngen case study. ArtificialTree is generated with the synthetic tree simulator in the PHATE package, which samples points from a branching hierarchy with Gaussian noise. The implementation is documented under \href{https://phate.readthedocs.io/en/stable/api.html#module-phate.tree}{\texttt{phate.tree.gen\_dla}}.

\begin{table}[H]
\centering
\scriptsize
\setlength{\tabcolsep}{2pt}
\renewcommand{\arraystretch}{1.04}
\caption{Datasets used in the nine-dataset benchmark and the \emph{C.\ elegans} case study. The separately generated \texttt{dyngen} simulation is described in Section~\ref{sec:supp-dyngen}. The table reports modality, sample size $n$, feature dimension $D$, number of class labels, and the annotation used for evaluation.\label{tab:datasets}}
\begin{tabular*}{\textwidth}{@{\extracolsep{\fill}}llrrrl@{\extracolsep{\fill}}}
\toprule
Dataset & Modality & $n$ & $D$ & Classes & Evaluation annotation\\
\midrule
ArtificialTree            & synthetic & 80{,}000      & 1{,}000 & 80 & Synthetic branch labels\\
MNIST                     & image     & 60{,}000      & 784     & 10 & Digit class ($0$--$9$)\\
EMNIST                    & image     & 697{,}932     & 784     & 62    & Character class\\
NG20                      & text      & 18{,}846      & 100     & 20 & Newsgroup topic\\
ACT                       & sensor    & 10{,}299      & 561     & 6  & Activity type\\
HCL                       & scRNA-seq & 49{,}551      & 3{,}038 & 57 & Human cell types across tissues\\
MCA                       & scRNA-seq & 23{,}341      & 9{,}120 & 19  & Mouse cell types\\
GAST10K                   & scRNA-seq & 10{,}638      & 1{,}458 & 12 & Human gastric-mucosa cell types\\
CELEGAN                   & scRNA-seq & 54{,}649      & 2{,}766 & 36 & \emph{C. elegans} embryonic cell types; developmental time\\
EPI                       & scRNA-seq & 100{,}000     & 500     & 10 & Epithelial cell/state\\
\botrule
\end{tabular*}
\end{table}

\section{Derivation of the dimension-free kNN log-density}
\label{sec:supp-density-derivation}
Let $r_k(\mathbf{x})$ denote the distance from a query observation $\mathbf{x}$ to its $k$-th nearest neighbor among $n$ reference observations in $d$ dimensions, excluding the query itself when it belongs to the reference sample. Assuming that the density $f$ is approximately constant within this neighborhood gives
\begin{equation}
\frac{k}{n}\approx f(\mathbf{x})\,V_d\,r_k(\mathbf{x})^{d},
\qquad
V_d=\frac{\pi^{d/2}}{\Gamma\!\left(\tfrac{d}{2}+1\right)},
\end{equation}
where $V_d$ is the volume of the unit $d$-ball. The corresponding kNN density estimator and its logarithm are
\begin{equation}
\hat{f}_k(\mathbf{x})
=\frac{k}{nV_d r_k(\mathbf{x})^d},
\qquad
\log\hat{f}_k(\mathbf{x})
=C-d\log r_k(\mathbf{x}),
\qquad C=\log k-\log n-\log V_d.
\end{equation}
For fixed $n$, $k$, and $d$, the term $C$ is constant across points and $d$ is a positive scale factor. Because Pearson correlation is invariant to positive affine transformations of either density vector, neither term affects the density-preservation objective. We therefore use the dimension-free quantity $\log\hat{f}_k(\mathbf{x})\propto-\log r_k(\mathbf{x})$.

\section{Model setup}
\label{sec:supp-model-setup}
The architecture and training settings reported here are the effective configurations used in the final five-seed benchmark. Input features are normalized with \texttt{BatchNorm1d} and compressed by a learned low-rank map into $M=32$ latent tokens with rank $r=16$. Each latent is independently expanded to token width $d_t=224$ and augmented with a learned latent-identity embedding. Two pre-LayerNorm Transformer blocks with four attention heads and feed-forward width $4d_t$ operate within the latent-token set of each observation. Both residual and attention dropout are zero. A final LayerNorm and mean pooling aggregate the tokens, after which a linear layer followed by batch normalization produces the 40-dimensional representation $\mathbf{z}_i=e_\phi(\mathbf{x}_i)$.
\subsection{Training and hyperparameter settings}
\label{sec:supp-hyperparameters}
For the main benchmark, we set the density-loss weight to $\lambda_d=1.8\times10^{-3}$, the neighborhood scale to $k=12$, the maximum number of density anchors to $A_0=512$, the rank-kernel base to $\eta=0.4$, the degrees of freedom of the embedding-space $t$ kernel to $\nu=0.01$, and the hard-pair threshold rank to $m_0=100$. We use $\varepsilon_{\mathrm{dist}}=10^{-12}$, $\varepsilon_{\mathrm{aff}}=10^{-6}$, $\varepsilon_{\mathrm{BCE}}=10^{-7}$, $\varepsilon_{\log}=10^{-8}$, and $\varepsilon_{\rho}=10^{-8}$ for numerical stability. Neighborhood augmentation uses $K_{\mathrm{aug}}=200$ and $D_{\mathrm{pca}}=64$, with independent interpolation weights $\alpha_i\sim\mathrm{Uniform}(0.05,1)$. We train the model for $1000$ epochs using AdamW with a learning rate of $1\times10^{-3}$, cosine annealing, a five-epoch warmup, mixed precision on one GPU, and a batch size of $4096$. Sensitivity to $\lambda_d$, $k$, and $A_0$ is reported in Section~\ref{sec:supp-param-sensitivity}.

\subsection{Baseline hyperparameter search}
\label{sec:supp-baseline-hp}
We perform hyperparameter selection in two stages. First, each baseline is evaluated on the two-axis structural grid in Table~\ref{tab:supp-baseline-hp}, yielding $5\times5=25$ configurations per method. Second, the selected structural configuration for den-SNE and densMAP is held fixed while the density-strength parameter is varied over $\lambda\in\{0.1,0.25,0.5,1,2,4\}$. At both stages, configurations are ranked by the arithmetic mean of density correlation and SVC accuracy.
Each selected configuration is then evaluated over five seeds for final reporting. Thus, class labels contribute to hyperparameter selection and evaluation but not to representation training. DMT-Dens settings are selected using the same combined criterion; sensitivity to its density-estimation parameters is reported in Section~\ref{sec:supp-param-sensitivity}.

\begin{table}[H]
\caption{Hyperparameter search grids for the baseline methods.\label{tab:supp-baseline-hp}}
\centering
\footnotesize
\setlength{\tabcolsep}{6pt}
\renewcommand{\arraystretch}{1.2}
\begin{tabular*}{\textwidth}{@{\extracolsep{\fill}}lll@{\extracolsep{\fill}}}
\toprule
Method & Hyperparameter & Search values \\
\midrule
t-SNE   & perplexity            & $15,\ 30,\ 50,\ 80,\ 120$ \\
        & early exaggeration    & $4,\ 8,\ 12,\ 18,\ 24$ \\
\noalign{\vskip3pt}
UMAP    & \texttt{n\_neighbors} & $10,\ 15,\ 20,\ 40,\ 80$ \\
        & \texttt{min\_dist}    & $0.001,\ 0.01,\ 0.05,\ 0.08,\ 0.15$ \\
\noalign{\vskip3pt}
den-SNE & perplexity            & $15,\ 30,\ 50,\ 80,\ 120$ \\
        & early exaggeration    & $4,\ 8,\ 12,\ 18,\ 24$ \\
        & density strength $\lambda$ (stage 2) & $0.1,\ 0.25,\ 0.5,\ 1,\ 2,\ 4$ \\
\noalign{\vskip3pt}
densMAP & \texttt{n\_neighbors} & $10,\ 15,\ 20,\ 40,\ 80$ \\
        & \texttt{min\_dist}    & $0.001,\ 0.01,\ 0.05,\ 0.08,\ 0.15$ \\
        & density strength $\lambda$ (stage 2) & $0.1,\ 0.25,\ 0.5,\ 1,\ 2,\ 4$ \\
\noalign{\vskip3pt}
PaCMAP  & \texttt{n\_neighbors} & $10,\ 15,\ 20,\ 40,\ 80$ \\
        & (\texttt{MN\_ratio}, \texttt{FP\_ratio}) & $(0.3,1),\ (0.5,1),\ (0.5,2),\ (1,2),\ (2,5)$ \\
\noalign{\vskip3pt}
PHATE   & \texttt{knn}          & $5,\ 10,\ 15,\ 20,\ 40$ \\
        & \texttt{decay}        & $10,\ 20,\ 40,\ 80,\ 120$ \\
\botrule
\end{tabular*}
\end{table}

\section{Parameter sensitivity}
\label{sec:supp-param-sensitivity}
We assess one-at-a-time sensitivity to three density-estimation hyperparameters: the maximum number of density anchors $A_0$, the neighborhood scale $k$, and the density-loss weight $\lambda_d$. In each table, the other two parameters are held at the main-benchmark setting ($A_0=512$, $k=12$, and $\lambda_d=1.8\times10^{-3}$). The central setting is represented by the same five baseline runs with seeds 42--46 in all three tables.

\begin{table}[H]
\caption{Sensitivity to the maximum number of density anchors $A_0$ on GAST10K, MCA, HCL, and NG20. Entries are mean$\pm$standard deviation over five seeds; higher is better for all three reported metrics.\label{tab:supp-param-anchor}}
\centering
\footnotesize
\setlength{\tabcolsep}{6pt}
\renewcommand{\arraystretch}{1.1}
\begin{tabular*}{\columnwidth}{@{}l@{\extracolsep{\fill}}lccc@{}}
\toprule
Dataset & Setting & Density & Local dens. & SVC \\
\midrule
GAST10K & $A_0=128$ & $0.821\pm0.022$ & $0.824\pm0.021$ & $0.794\pm0.045$ \\
 & $A_0=256$ & $0.853\pm0.013$ & $0.848\pm0.013$ & $0.768\pm0.054$ \\
 & $A_0=512$ & $0.873\pm0.009$ & $0.856\pm0.011$ & $0.787\pm0.051$ \\
 & $A_0=768$ & $0.871\pm0.013$ & $0.848\pm0.014$ & $0.765\pm0.037$ \\
 & $A_0=1024$ & $0.878\pm0.012$ & $0.850\pm0.015$ & $0.727\pm0.067$ \\
 & $A_0=1536$ & $0.884\pm0.010$ & $0.852\pm0.015$ & $0.789\pm0.043$ \\
\midrule
MCA & $A_0=128$ & $0.661\pm0.040$ & $0.612\pm0.043$ & $0.739\pm0.083$ \\
 & $A_0=256$ & $0.648\pm0.044$ & $0.581\pm0.040$ & $0.804\pm0.029$ \\
 & $A_0=512$ & $0.642\pm0.063$ & $0.551\pm0.086$ & $0.788\pm0.026$ \\
 & $A_0=768$ & $0.612\pm0.071$ & $0.529\pm0.069$ & $0.752\pm0.045$ \\
 & $A_0=1024$ & $0.608\pm0.030$ & $0.516\pm0.045$ & $0.750\pm0.042$ \\
 & $A_0=1536$ & $0.597\pm0.014$ & $0.504\pm0.031$ & $0.783\pm0.084$ \\
\midrule
HCL & $A_0=128$ & $0.837\pm0.011$ & $0.818\pm0.018$ & $0.774\pm0.013$ \\
 & $A_0=256$ & $0.847\pm0.010$ & $0.819\pm0.019$ & $0.778\pm0.008$ \\
 & $A_0=512$ & $0.864\pm0.005$ & $0.829\pm0.011$ & $0.763\pm0.008$ \\
 & $A_0=768$ & $0.867\pm0.011$ & $0.826\pm0.018$ & $0.774\pm0.016$ \\
 & $A_0=1024$ & $0.871\pm0.011$ & $0.831\pm0.018$ & $0.764\pm0.020$ \\
 & $A_0=1536$ & $0.871\pm0.015$ & $0.821\pm0.019$ & $0.769\pm0.031$ \\
\midrule
NG20 & $A_0=128$ & $0.578\pm0.011$ & $0.589\pm0.005$ & $0.335\pm0.010$ \\
 & $A_0=256$ & $0.649\pm0.020$ & $0.650\pm0.018$ & $0.345\pm0.026$ \\
 & $A_0=512$ & $0.678\pm0.023$ & $0.676\pm0.023$ & $0.334\pm0.027$ \\
 & $A_0=768$ & $0.714\pm0.035$ & $0.700\pm0.026$ & $0.331\pm0.037$ \\
 & $A_0=1024$ & $0.723\pm0.031$ & $0.702\pm0.024$ & $0.334\pm0.044$ \\
 & $A_0=1536$ & $0.759\pm0.026$ & $0.728\pm0.021$ & $0.307\pm0.019$ \\
\botrule
\end{tabular*}
\end{table}

\begin{table}[H]
\caption{Sensitivity to the density neighborhood scale $k$ on GAST10K, MCA, HCL, and NG20. Entries are mean$\pm$standard deviation over five seeds; higher is better for all three reported metrics.\label{tab:supp-param-k}}
\centering
\footnotesize
\setlength{\tabcolsep}{6pt}
\renewcommand{\arraystretch}{1.1}
\begin{tabular*}{\columnwidth}{@{}l@{\extracolsep{\fill}}lccc@{}}
\toprule
Dataset & Setting & Density & Local dens. & SVC \\
\midrule
GAST10K & $k=5$ & $0.809\pm0.021$ & $0.796\pm0.024$ & $0.792\pm0.045$ \\
 & $k=10$ & $0.859\pm0.012$ & $0.838\pm0.010$ & $0.749\pm0.033$ \\
 & $k=12$ & $0.873\pm0.009$ & $0.856\pm0.011$ & $0.787\pm0.051$ \\
 & $k=15$ & $0.868\pm0.009$ & $0.860\pm0.013$ & $0.728\pm0.054$ \\
 & $k=20$ & $0.871\pm0.011$ & $0.876\pm0.010$ & $0.782\pm0.058$ \\
 & $k=25$ & $0.863\pm0.005$ & $0.881\pm0.005$ & $0.760\pm0.072$ \\
 & $k=30$ & $0.852\pm0.008$ & $0.885\pm0.010$ & $0.781\pm0.028$ \\
 & $k=40$ & $0.800\pm0.013$ & $0.874\pm0.010$ & $0.765\pm0.050$ \\
\midrule
MCA & $k=5$ & $0.566\pm0.065$ & $0.496\pm0.076$ & $0.747\pm0.079$ \\
 & $k=10$ & $0.639\pm0.062$ & $0.535\pm0.069$ & $0.793\pm0.026$ \\
 & $k=12$ & $0.642\pm0.063$ & $0.551\pm0.086$ & $0.788\pm0.026$ \\
 & $k=15$ & $0.612\pm0.037$ & $0.559\pm0.033$ & $0.764\pm0.045$ \\
 & $k=20$ & $0.647\pm0.084$ & $0.621\pm0.082$ & $0.792\pm0.010$ \\
 & $k=25$ & $0.666\pm0.044$ & $0.647\pm0.042$ & $0.786\pm0.044$ \\
 & $k=30$ & $0.596\pm0.056$ & $0.592\pm0.059$ & $0.749\pm0.070$ \\
 & $k=40$ & $0.514\pm0.060$ & $0.516\pm0.062$ & $0.770\pm0.030$ \\
\midrule
HCL & $k=5$ & $0.797\pm0.061$ & $0.767\pm0.068$ & $0.755\pm0.019$ \\
 & $k=10$ & $0.853\pm0.011$ & $0.817\pm0.019$ & $0.768\pm0.021$ \\
 & $k=12$ & $0.864\pm0.005$ & $0.829\pm0.011$ & $0.763\pm0.008$ \\
 & $k=15$ & $0.865\pm0.009$ & $0.838\pm0.015$ & $0.761\pm0.014$ \\
 & $k=20$ & $0.869\pm0.005$ & $0.860\pm0.009$ & $0.781\pm0.021$ \\
 & $k=25$ & $0.859\pm0.009$ & $0.859\pm0.014$ & $0.791\pm0.016$ \\
 & $k=30$ & $0.841\pm0.012$ & $0.854\pm0.010$ & $0.779\pm0.026$ \\
 & $k=40$ & $0.792\pm0.011$ & $0.833\pm0.012$ & $0.791\pm0.017$ \\
\midrule
NG20 & $k=5$ & $0.526\pm0.019$ & $0.526\pm0.017$ & $0.327\pm0.034$ \\
 & $k=10$ & $0.671\pm0.036$ & $0.666\pm0.035$ & $0.297\pm0.035$ \\
 & $k=12$ & $0.678\pm0.023$ & $0.676\pm0.023$ & $0.334\pm0.027$ \\
 & $k=15$ & $0.680\pm0.020$ & $0.683\pm0.018$ & $0.318\pm0.041$ \\
 & $k=20$ & $0.709\pm0.017$ & $0.718\pm0.018$ & $0.320\pm0.024$ \\
 & $k=25$ & $0.694\pm0.011$ & $0.713\pm0.014$ & $0.308\pm0.036$ \\
 & $k=30$ & $0.689\pm0.011$ & $0.727\pm0.010$ & $0.315\pm0.040$ \\
 & $k=40$ & $0.660\pm0.009$ & $0.726\pm0.013$ & $0.329\pm0.010$ \\
\botrule
\end{tabular*}
\end{table}

\begin{table}[H]
\caption{Sensitivity to the density loss weight $\lambda_d$ on GAST10K, MCA, HCL, and NG20. Entries are mean$\pm$standard deviation over five seeds; higher is better for all three reported metrics.\label{tab:supp-param-lambda}}
\centering
\footnotesize
\setlength{\tabcolsep}{6pt}
\renewcommand{\arraystretch}{1.1}
\begin{tabular*}{\columnwidth}{@{}l@{\extracolsep{\fill}}lccc@{}}
\toprule
Dataset & Setting & Density & Local dens. & SVC \\
\midrule
GAST10K & $\lambda_d=0.0001$ & $0.223\pm0.063$ & $0.186\pm0.068$ & $0.726\pm0.062$ \\
 & $\lambda_d=0.0005$ & $0.742\pm0.020$ & $0.665\pm0.030$ & $0.752\pm0.049$ \\
 & $\lambda_d=0.001$ & $0.837\pm0.007$ & $0.803\pm0.014$ & $0.753\pm0.069$ \\
 & $\lambda_d=0.0018$ & $0.873\pm0.009$ & $0.856\pm0.011$ & $0.787\pm0.051$ \\
 & $\lambda_d=0.003$ & $0.883\pm0.002$ & $0.876\pm0.006$ & $0.760\pm0.042$ \\
 & $\lambda_d=0.006$ & $0.907\pm0.004$ & $0.904\pm0.007$ & $0.767\pm0.045$ \\
 & $\lambda_d=0.01$ & $0.920\pm0.007$ & $0.920\pm0.008$ & $0.746\pm0.044$ \\
 & $\lambda_d=0.02$ & $0.936\pm0.006$ & $0.937\pm0.005$ & $0.724\pm0.050$ \\
\midrule
MCA & $\lambda_d=0.0001$ & $0.323\pm0.051$ & $0.239\pm0.061$ & $0.790\pm0.065$ \\
 & $\lambda_d=0.0005$ & $0.434\pm0.040$ & $0.359\pm0.040$ & $0.821\pm0.030$ \\
 & $\lambda_d=0.001$ & $0.564\pm0.060$ & $0.489\pm0.057$ & $0.813\pm0.017$ \\
 & $\lambda_d=0.0018$ & $0.642\pm0.063$ & $0.551\pm0.086$ & $0.788\pm0.026$ \\
 & $\lambda_d=0.003$ & $0.719\pm0.078$ & $0.646\pm0.083$ & $0.786\pm0.044$ \\
 & $\lambda_d=0.006$ & $0.778\pm0.019$ & $0.728\pm0.031$ & $0.758\pm0.052$ \\
 & $\lambda_d=0.01$ & $0.834\pm0.019$ & $0.802\pm0.026$ & $0.689\pm0.027$ \\
 & $\lambda_d=0.02$ & $0.831\pm0.024$ & $0.809\pm0.029$ & $0.657\pm0.038$ \\
\midrule
HCL & $\lambda_d=0.0001$ & $0.441\pm0.025$ & $0.420\pm0.024$ & $0.825\pm0.016$ \\
 & $\lambda_d=0.0005$ & $0.751\pm0.007$ & $0.703\pm0.014$ & $0.804\pm0.018$ \\
 & $\lambda_d=0.001$ & $0.819\pm0.010$ & $0.769\pm0.023$ & $0.796\pm0.014$ \\
 & $\lambda_d=0.0018$ & $0.864\pm0.005$ & $0.829\pm0.011$ & $0.763\pm0.008$ \\
 & $\lambda_d=0.003$ & $0.886\pm0.008$ & $0.861\pm0.014$ & $0.750\pm0.020$ \\
 & $\lambda_d=0.006$ & $0.914\pm0.004$ & $0.898\pm0.004$ & $0.711\pm0.028$ \\
 & $\lambda_d=0.01$ & $0.924\pm0.003$ & $0.915\pm0.005$ & $0.695\pm0.015$ \\
 & $\lambda_d=0.02$ & $0.934\pm0.003$ & $0.930\pm0.002$ & $0.625\pm0.015$ \\
\midrule
NG20 & $\lambda_d=0.0001$ & $0.196\pm0.026$ & $0.229\pm0.021$ & $0.331\pm0.022$ \\
 & $\lambda_d=0.0005$ & $0.565\pm0.038$ & $0.539\pm0.030$ & $0.322\pm0.008$ \\
 & $\lambda_d=0.001$ & $0.646\pm0.022$ & $0.625\pm0.021$ & $0.319\pm0.036$ \\
 & $\lambda_d=0.0018$ & $0.678\pm0.023$ & $0.676\pm0.023$ & $0.334\pm0.027$ \\
 & $\lambda_d=0.003$ & $0.712\pm0.019$ & $0.710\pm0.020$ & $0.301\pm0.054$ \\
 & $\lambda_d=0.006$ & $0.762\pm0.016$ & $0.766\pm0.012$ & $0.286\pm0.035$ \\
 & $\lambda_d=0.01$ & $0.812\pm0.008$ & $0.819\pm0.009$ & $0.259\pm0.013$ \\
 & $\lambda_d=0.02$ & $0.895\pm0.015$ & $0.899\pm0.017$ & $0.183\pm0.022$ \\
\botrule
\end{tabular*}
\end{table}

\section{Runtime, peak GPU memory, and hardware}
Matched runtime and peak-memory measurements were collected on a workstation with two Intel Xeon Gold 5118 CPUs at 2.30\,GHz (24 physical cores and 48 threads in total), 376\,GiB of system memory, and eight NVIDIA GeForce RTX 2080 Ti GPUs with 11\,GB of GDDR6 memory each. The system ran Ubuntu 22.04.5 LTS with NVIDIA driver 545.23.08, for which \texttt{nvidia-smi} reported CUDA 12.3 compatibility.
DMT-Dens was implemented in Python 3.10.19 using PyTorch 2.5.1+cu121, Lightning 2.5.4, and cuDNN 9.1.0; the PyTorch build used CUDA 12.1. Baselines were implemented with openTSNE 1.0.4 for t-SNE, \texttt{umap-learn} 0.5.9.post2 for UMAP and densMAP, PaCMAP 0.8.0, PHATE 2.0.0, and the repository-bundled 2020 implementation of den-SNE. Evaluation used \texttt{scikit-learn} 1.7.2. Each DMT-Dens run used one RTX 2080 Ti GPU, whereas the non-parametric baselines (t-SNE, UMAP, den-SNE, densMAP, PHATE, and PaCMAP) were run on CPU.

The runtime analysis reports wall-clock fitting time from the matched runtime sweep. Values are the mean$\pm$standard deviation over three seeds, measured in seconds. Peak GPU memory under the full training configuration is reported in Table~\ref{tab:supp-memory-all}.

\begin{table}[H]
\caption{Runtime scaling with dataset size. Entries are wall-clock fitting times in seconds, reported as mean$\pm$standard deviation over three seeds.
\label{tab:supp-runtime}}
\centering
\footnotesize
\setlength{\tabcolsep}{3pt}
\renewcommand{\arraystretch}{1.05}
\begin{tabular*}{\textwidth}{@{\extracolsep{\fill}}llcccccc@{\extracolsep{\fill}}}
\toprule
Dataset & $n$ & t-SNE & UMAP & den-SNE & densMAP & PHATE & DMT-Dens \\
\midrule
MNIST & 2,000 & $12.9\pm0.2$ & $35.6\pm0.3$ & $21.8\pm0.4$ & $29.8\pm0.2$ & $16.5\pm0.5$ & $119.2\pm12.4$ \\
 & 5,000 & $25.9\pm0.5$ & $56.4\pm0.8$ & $61.8\pm0.9$ & $47.7\pm8.5$ & $30.5\pm0.6$ & $152.7\pm6.4$ \\
 & 10,000 & $82.7\pm1.6$ & $71.7\pm1.2$ & $141.8\pm9.8$ & $72.2\pm8.5$ & $35.1\pm1.0$ & $232.9\pm64.7$ \\
 & 20,000 & $108.2\pm1.2$ & $71.9\pm1.3$ & $376.9\pm14.2$ & $78.7\pm1.3$ & $48.1\pm0.7$ & $321.0\pm3.9$ \\
 & 40,000 & $146.3\pm5.8$ & $105.6\pm1.4$ & $891.7\pm42.2$ & $152.4\pm5.3$ & $70.4\pm0.2$ & $427.2\pm39.6$ \\
 & 60,000 & $176.9\pm3.4$ & $138.2\pm1.9$ & $1635.9\pm27.8$ & $213.4\pm7.4$ & $102.7\pm4.1$ & $508.5\pm12.7$ \\
\midrule
EMNIST & 2,000 & $12.7\pm0.2$ & $35.0\pm0.2$ & $26.7\pm0.0$ & $29.9\pm0.2$ & $15.8\pm0.0$ & $40.4\pm1.1$ \\
 & 5,000 & $27.3\pm1.1$ & $56.8\pm0.9$ & $75.7\pm1.4$ & $56.0\pm0.3$ & $29.6\pm0.6$ & $73.0\pm3.0$ \\
 & 10,000 & $82.2\pm0.7$ & $72.0\pm1.5$ & $134.0\pm0.7$ & $78.9\pm2.0$ & $34.8\pm1.4$ & $98.5\pm1.9$ \\
 & 20,000 & $112.6\pm3.8$ & $73.2\pm0.5$ & $319.1\pm1.4$ & $96.2\pm0.8$ & $45.1\pm0.7$ & $155.3\pm4.8$ \\
 & 40,000 & $148.3\pm1.5$ & $104.2\pm0.9$ & $976.5\pm7.2$ & $159.9\pm1.8$ & $70.1\pm0.6$ & $307.8\pm2.5$ \\
 & 60,000 & $179.1\pm3.4$ & $139.1\pm0.5$ & $1746.3\pm1.5$ & $233.7\pm2.1$ & $104.0\pm1.0$ & $460.4\pm2.1$ \\
\midrule
GAST10K & 2,000 & $19.2\pm0.5$ & $42.9\pm1.5$ & $29.4\pm0.4$ & $37.6\pm1.0$ & $20.8\pm1.0$ & $89.1\pm7.6$ \\
 & 5,000 & $38.1\pm2.2$ & $72.3\pm1.7$ & $73.9\pm1.7$ & $71.8\pm0.3$ & $39.5\pm1.0$ & $127.4\pm5.0$ \\
 & 10,000 & $52.3\pm1.6$ & $97.9\pm1.2$ & $138.9\pm12.5$ & $97.8\pm0.5$ & $46.2\pm1.4$ & $169.3\pm11.1$ \\
 & 10,638 & $53.7\pm1.7$ & $87.7\pm0.2$ & $157.6\pm8.0$ & $85.8\pm0.8$ & $45.9\pm0.7$ & $185.0\pm9.7$ \\
\midrule
HCL & 2,000 & $31.7\pm1.2$ & $60.5\pm2.5$ & $50.6\pm1.7$ & $51.5\pm1.5$ & $33.2\pm1.4$ & $107.6\pm6.6$ \\
 & 5,000 & $65.1\pm3.3$ & $96.5\pm1.8$ & $105.1\pm3.7$ & $93.1\pm2.4$ & $60.1\pm1.7$ & $173.9\pm12.0$ \\
 & 10,000 & $87.9\pm3.1$ & $123.9\pm3.2$ & $191.9\pm7.4$ & $124.1\pm1.9$ & $65.1\pm1.6$ & $217.2\pm49.8$ \\
 & 20,000 & $128.6\pm0.2$ & $170.1\pm5.3$ & $403.7\pm24.5$ & $156.5\pm1.3$ & $79.9\pm0.7$ & $399.4\pm78.5$ \\
 & 40,000 & $184.5\pm6.9$ & $328.4\pm6.5$ & $1004.4\pm43.6$ & $261.1\pm3.4$ & $119.9\pm1.0$ & $636.0\pm8.3$ \\
 & 49,551 & $215.3\pm8.6$ & $430.8\pm15.7$ & $1371.0\pm84.6$ & $312.5\pm8.5$ & $139.0\pm1.9$ & $852.7\pm149.7$ \\
\midrule
MCA & 2,000 & $79.2\pm1.3$ & $123.7\pm2.9$ & $358.2\pm10.8$ & $101.2\pm2.2$ & $76.8\pm2.5$ & $210.3\pm103.6$ \\
 & 5,000 & $163.8\pm0.5$ & $189.4\pm4.1$ & $451.4\pm13.0$ & $177.1\pm2.9$ & $132.1\pm3.1$ & $293.0\pm48.9$ \\
 & 10,000 & $199.8\pm4.0$ & $251.7\pm5.1$ & $579.2\pm25.1$ & $214.8\pm0.9$ & $141.7\pm3.1$ & $392.6\pm21.1$ \\
 & 20,000 & $362.5\pm2.9$ & $369.3\pm7.8$ & $804.6\pm37.1$ & $274.3\pm3.1$ & $158.2\pm2.6$ & $459.4\pm31.2$ \\
 & 23,341 & $320.9\pm5.1$ & $424.6\pm10.9$ & $915.6\pm50.1$ & $301.8\pm6.9$ & $163.3\pm2.3$ & $511.4\pm13.2$ \\
\botrule
\end{tabular*}
\end{table}

Table~\ref{tab:supp-memory-all} reports peak GPU memory for the nine benchmark datasets, defined as the maximum CUDA memory allocated during training. Peak usage does not exceed approximately $6.7$\,GB.

\begin{table}[H]
\caption{Peak GPU memory of DMT-Dens under the full training configuration for the nine benchmark datasets. Values are the maximum CUDA memory allocated during training.
\label{tab:supp-memory-all}}
\centering
\footnotesize
\setlength{\tabcolsep}{14pt}
\renewcommand{\arraystretch}{1.1}
\begin{tabular}{lr}
\toprule
Dataset & Peak GPU mem.\ (MB) \\
\midrule
ACT & 6032.2 \\
EMNIST & 5923.4 \\
EPI & 5898.7 \\
GAST10K & 5982.4 \\
HCL & 6120.4 \\
MCA & 6649.8 \\
MNIST & 5923.4 \\
NG20 & 5864.8 \\
ArtificialTree & 5943.2 \\
\botrule
\end{tabular}
\end{table}

\section{Full quantitative metrics}
\label{sec:supp-full-metrics}
\label{sec:supp-local-density}
This section reports local-structure metrics that complement the main-text summaries of density correlation and SVC accuracy (Table~\ref{tab:density-local-svc-main}). Table~\ref{tab:supp-local-density} reports local density correlation and scattered point intrusion rate (SPIR), and Table~\ref{tab:supplementary-metrics} reports continuity and kNN preservation.

Density correlation compares pointwise kNN-radius estimates over the full dataset and thus summarizes the overall ordering of dense and sparse regions. Local density correlation evaluates density agreement within local neighborhoods, making it more sensitive to short-range density variation. SPIR quantifies spurious neighborhood intrusions in the embedding, in which points that are distant in the input space become locally close after projection; lower values indicate fewer such intrusions.

\begin{table}[H]
\caption{Density-preservation metrics for DMT-Dens and six baselines across the benchmark datasets. Higher local density correlation and lower scattered point intrusion rate (SPIR) are better. The best value in each row is shown in \textbf{bold}. Entries are mean$\pm$standard deviation over the available seeds. The final row of each panel reports the unweighted mean over all nine datasets; this mean is omitted for den-SNE because its EMNIST run did not complete. OOT (out of time) denotes a runtime of 24\,h+ for den-SNE on the full EMNIST dataset.\label{tab:supp-local-density}}
\centering
\scriptsize
\setlength{\tabcolsep}{2pt}
\renewcommand{\arraystretch}{1.0}

\textbf{\textit{Density-preservation metrics}}\par\vspace{2pt}
\textbf{Local density correlation}\par\vspace{2pt}
\begin{tabular*}{\textwidth}{@{\extracolsep{\fill}}lccccccc@{\extracolsep{\fill}}}
\toprule
Dataset & t-SNE & UMAP & den-SNE & densMAP & PHATE & PaCMAP & DMT-Dens \\
\midrule
ArtificialTree & $0.791\pm0.015$ & $0.566\pm0.047$ & $0.535\pm0.034$ & $0.638\pm0.058$ & $-0.421\pm0.082$ & $0.685\pm0.031$ & \best{$0.904\pm0.011$} \\
ACT    & $0.408\pm0.061$ & $0.220\pm0.052$ & \best{$0.908\pm0.009$} & $0.866\pm0.011$ & $0.449\pm0.098$ & $0.159\pm0.042$ & $0.839\pm0.006$ \\
EMNIST & $0.228\pm0.035$ & $0.045\pm0.024$ & OOT & $0.605\pm0.025$ & $0.167\pm0.055$ & $0.097\pm0.020$ & \best{$0.673\pm0.020$} \\
MNIST  & $0.262\pm0.046$ & $-0.110\pm0.028$ & \best{$0.796\pm0.015$} & $0.781\pm0.021$ & $0.169\pm0.021$ & $-0.042\pm0.044$ & $0.712\pm0.035$ \\
NG20   & $0.446\pm0.032$ & $0.350\pm0.022$ & $0.737\pm0.008$ & \best{$0.830\pm0.008$} & $0.665\pm0.013$ & $0.259\pm0.044$ & $0.666\pm0.019$ \\
EPI    & $0.226\pm0.033$ & $0.103\pm0.032$ & $0.697\pm0.012$ & $0.716\pm0.010$ & $0.251\pm0.015$ & $0.143\pm0.024$ & \best{$0.774\pm0.018$} \\
GAST10K & $0.059\pm0.040$ & $-0.164\pm0.021$ & $0.793\pm0.016$ & $0.810\pm0.015$ & $0.187\pm0.017$ & $-0.526\pm0.028$ & \best{$0.858\pm0.010$} \\
HCL    & $0.120\pm0.025$ & $0.041\pm0.039$ & $0.716\pm0.016$ & $0.542\pm0.020$ & $0.407\pm0.010$ & $0.142\pm0.038$ & \best{$0.838\pm0.004$} \\
MCA    & $0.389\pm0.022$ & $0.096\pm0.016$ & $0.301\pm0.015$ & $0.493\pm0.022$ & $-0.029\pm0.032$ & $0.004\pm0.033$ & \best{$0.557\pm0.068$} \\
\midrule
Mean   & $0.325$ & $0.127$ & -- & $0.698$ & $0.205$ & $0.102$ & \best{$0.758$} \\
\botrule
\end{tabular*}

\vspace{3pt}
\textbf{SPIR} (lower is better)\par\vspace{2pt}
\begin{tabular*}{\textwidth}{@{\extracolsep{\fill}}lccccccc@{\extracolsep{\fill}}}
\toprule
Dataset & t-SNE & UMAP & den-SNE & densMAP & PHATE & PaCMAP & DMT-Dens \\
\midrule
ArtificialTree & $0.615\pm0.031$ & $0.637\pm0.032$ & $0.634\pm0.030$ & $0.639\pm0.021$ & \best{$0.604\pm0.036$} & $0.626\pm0.016$ & $0.619\pm0.026$ \\
ACT    & $0.448\pm0.016$ & $0.472\pm0.024$ & $0.435\pm0.005$ & $0.494\pm0.019$ & $0.463\pm0.052$ & $0.378\pm0.013$ & \best{$0.362\pm0.011$} \\
EMNIST & $0.733\pm0.019$ & $0.744\pm0.024$ & OOT & $0.768\pm0.020$ & $0.802\pm0.000$ & $0.703\pm0.020$ & \best{$0.674\pm0.019$} \\
MNIST  & $0.670\pm0.019$ & $0.702\pm0.007$ & $0.649\pm0.028$ & $0.705\pm0.015$ & $0.748\pm0.011$ & $0.647\pm0.018$ & \best{$0.624\pm0.032$} \\
NG20   & $0.608\pm0.028$ & $0.564\pm0.016$ & $0.621\pm0.026$ & \best{$0.505\pm0.013$} & $0.560\pm0.020$ & $0.584\pm0.012$ & $0.660\pm0.032$ \\
EPI    & $0.778\pm0.015$ & $0.829\pm0.017$ & $0.688\pm0.017$ & $0.668\pm0.020$ & $0.794\pm0.014$ & $0.632\pm0.027$ & \best{$0.612\pm0.033$} \\
GAST10K & $0.341\pm0.020$ & $0.393\pm0.023$ & $0.207\pm0.026$ & \best{$0.162\pm0.025$} & $0.214\pm0.022$ & $0.180\pm0.013$ & $0.215\pm0.022$ \\
HCL    & $0.691\pm0.006$ & $0.720\pm0.026$ & $0.590\pm0.022$ & $0.681\pm0.027$ & $0.650\pm0.017$ & $0.590\pm0.016$ & \best{$0.573\pm0.016$} \\
MCA    & $0.740\pm0.024$ & $0.789\pm0.012$ & $0.656\pm0.019$ & $0.764\pm0.019$ & $0.695\pm0.024$ & \best{$0.654\pm0.013$} & $0.658\pm0.012$ \\
\midrule
Mean   & $0.625$ & $0.650$ & -- & $0.598$ & $0.614$ & \best{$0.555$} & $0.555$ \\
\botrule
\end{tabular*}
\end{table}

\begin{table}[H]
\caption{Neighbor-ranking metrics for DMT-Dens and six baselines across the benchmark datasets. Higher continuity and kNN preservation are better, and the best value in each row is shown in \textbf{bold}. Seed averaging, unweighted means, and the OOT marker follow Table~\ref{tab:supp-local-density}.\label{tab:supplementary-metrics}}
\centering
\scriptsize
\setlength{\tabcolsep}{2pt}
\renewcommand{\arraystretch}{1.0}

\textbf{\textit{Neighbor-ranking metrics}}\par\vspace{2pt}
\textbf{Continuity}\par\vspace{2pt}
\begin{tabular*}{\textwidth}{@{\extracolsep{\fill}}lccccccc@{\extracolsep{\fill}}}
\toprule
Dataset & t-SNE & UMAP & den-SNE & densMAP & PHATE & PaCMAP & DMT-Dens \\
\midrule
ArtificialTree & $0.997\pm0.000$ & $0.997\pm0.000$ & $0.986\pm0.001$ & \best{$0.998\pm0.000$} & $0.958\pm0.035$ & $0.992\pm0.001$ & $0.996\pm0.001$ \\
ACT    & $0.972\pm0.000$ & $0.969\pm0.002$ & $0.971\pm0.001$ & $0.970\pm0.001$ & $0.967\pm0.002$ & \best{$0.975\pm0.001$} & $0.973\pm0.001$ \\
EMNIST & $0.898\pm0.003$ & $0.887\pm0.004$ & OOT & $0.892\pm0.003$ & \best{$0.910\pm0.001$} & $0.889\pm0.002$ & $0.882\pm0.009$ \\
MNIST  & $0.948\pm0.002$ & $0.947\pm0.002$ & \best{$0.949\pm0.002$} & $0.949\pm0.002$ & $0.949\pm0.003$ & $0.939\pm0.003$ & $0.941\pm0.003$ \\
NG20   & $0.879\pm0.009$ & $0.884\pm0.006$ & $0.891\pm0.002$ & \best{$0.912\pm0.002$} & $0.892\pm0.007$ & $0.888\pm0.003$ & $0.864\pm0.004$ \\
EPI    & $0.929\pm0.006$ & $0.929\pm0.006$ & $0.917\pm0.008$ & \best{$0.932\pm0.006$} & $0.931\pm0.005$ & $0.920\pm0.004$ & $0.921\pm0.007$ \\
GAST10K & \best{$0.831\pm0.011$} & $0.826\pm0.006$ & $0.762\pm0.006$ & $0.827\pm0.006$ & $0.804\pm0.006$ & $0.739\pm0.003$ & $0.761\pm0.009$ \\
HCL    & \best{$0.797\pm0.011$} & $0.697\pm0.021$ & $0.737\pm0.016$ & $0.710\pm0.008$ & $0.792\pm0.002$ & $0.753\pm0.004$ & $0.759\pm0.008$ \\
MCA    & $0.740\pm0.023$ & $0.799\pm0.015$ & $0.607\pm0.025$ & \best{$0.825\pm0.006$} & $0.617\pm0.027$ & $0.666\pm0.027$ & $0.751\pm0.056$ \\
\midrule
Mean   & $0.888$ & $0.882$ & -- & \best{$0.891$} & $0.869$ & $0.862$ & $0.872$ \\
\botrule
\end{tabular*}

\vspace{3pt}
\textbf{kNN preservation}\par\vspace{2pt}
\begin{tabular*}{\textwidth}{@{\extracolsep{\fill}}lccccccc@{\extracolsep{\fill}}}
\toprule
Dataset & t-SNE & UMAP & den-SNE & densMAP & PHATE & PaCMAP & DMT-Dens \\
\midrule
ArtificialTree & \best{$0.913\pm0.003$} & $0.708\pm0.008$ & $0.783\pm0.008$ & $0.752\pm0.005$ & $0.289\pm0.066$ & $0.833\pm0.006$ & $0.900\pm0.005$ \\
ACT    & \best{$0.423\pm0.003$} & $0.370\pm0.004$ & $0.392\pm0.004$ & $0.298\pm0.004$ & $0.250\pm0.014$ & $0.300\pm0.005$ & $0.297\pm0.007$ \\
EMNIST & \best{$0.347\pm0.005$} & $0.320\pm0.005$ & OOT & $0.302\pm0.008$ & $0.168\pm0.003$ & $0.328\pm0.005$ & $0.312\pm0.004$ \\
MNIST  & \best{$0.371\pm0.002$} & $0.318\pm0.004$ & $0.322\pm0.002$ & $0.246\pm0.006$ & $0.178\pm0.007$ & $0.325\pm0.005$ & $0.315\pm0.003$ \\
NG20   & \best{$0.337\pm0.010$} & $0.297\pm0.005$ & $0.274\pm0.005$ & $0.210\pm0.007$ & $0.142\pm0.008$ & $0.303\pm0.009$ & $0.257\pm0.004$ \\
EPI    & \best{$0.181\pm0.004$} & $0.140\pm0.003$ & $0.157\pm0.004$ & $0.128\pm0.003$ & $0.115\pm0.004$ & $0.146\pm0.004$ & $0.154\pm0.007$ \\
GAST10K & \best{$0.083\pm0.003$} & $0.068\pm0.004$ & $0.076\pm0.003$ & $0.042\pm0.001$ & $0.042\pm0.002$ & $0.049\pm0.002$ & $0.056\pm0.001$ \\
HCL    & \best{$0.117\pm0.005$} & $0.080\pm0.003$ & $0.098\pm0.005$ & $0.077\pm0.003$ & $0.062\pm0.003$ & $0.090\pm0.005$ & $0.088\pm0.007$ \\
MCA    & \best{$0.097\pm0.005$} & $0.055\pm0.002$ & $0.092\pm0.007$ & $0.067\pm0.004$ & $0.064\pm0.004$ & $0.080\pm0.006$ & $0.084\pm0.004$ \\
\midrule
Mean   & \best{$0.319$} & $0.262$ & -- & $0.236$ & $0.146$ & $0.273$ & $0.274$ \\
\botrule
\end{tabular*}
\end{table}

\section{Additional evidence}\label{sec:supp-addl}
We complement the main benchmark with two evaluations based on external reference information: synthetic \texttt{dyngen} trajectories with a known backbone and a \emph{C.\ elegans} embryonic time course with observed developmental-time annotations.

\subsection{Synthetic trajectories (\texttt{dyngen})}\label{sec:supp-dyngen}
We evaluate DMT-Dens on synthetic single-cell data generated with \texttt{dyngen} \citep{cannoodt2021dyngen}, which simulates cells along a prescribed trajectory backbone governed by a controlled gene-regulatory network. The resulting topology, branch assignments, and simulation times provide reference information for evaluating the recovered branching structure.

\noindent\textbf{Simulation setup.} We generate single-cell data along a tree-structured, multifurcating trajectory backbone. Each simulated cell has a reference branch assignment and simulation time.

\noindent\textbf{Evaluation.} We report the high-dimensional-to-embedding density correlation and local density correlation used in the main benchmark. Separately, the dyngen branch assignments and trajectory backbone are used to assess branch-label separability with a support-vector classifier (Branch SVC) and topology fidelity, defined as the $F_1$ agreement between the embedding neighborhood graph and the reference backbone. We compare DMT-Dens with the same six baselines used in the main text. Each method is subsequently evaluated over five seeds using its selected configuration.

\begin{table}[H]
\caption{Quantitative comparison on synthetic \texttt{dyngen} trajectories with a known reference topology. Higher values are better, and the best value in each column is shown in \textbf{bold}. Each reported configuration maximizes $(\mathrm{density\ correlation}+\mathrm{Branch\ SVC})/2$ over the corresponding hyperparameter sweep. Entries are mean$\pm$standard deviation over five seeds.\label{tab:supp-dyngen}}
\centering
\scriptsize
\setlength{\tabcolsep}{3pt}
\renewcommand{\arraystretch}{1.04}
\begin{tabular*}{\columnwidth}{@{\extracolsep{\fill}}lcccc@{}}
\toprule
Method & Density corr. & Local density corr. & Branch SVC & Topology fidelity \\
\midrule
t-SNE      & $0.064\pm0.022$ & $0.057\pm0.019$ & \best{$0.922\pm0.003$} & $0.885\pm0.000$ \\
UMAP       & $0.267\pm0.018$ & $0.442\pm0.015$ & $0.888\pm0.033$ & $0.854\pm0.042$ \\
den-SNE    & $0.222\pm0.027$ & $0.259\pm0.021$ & $0.915\pm0.018$ & \best{$0.923\pm0.000$} \\
densMAP    & $0.754\pm0.009$ & $0.801\pm0.003$ & $0.881\pm0.009$ & $0.877\pm0.032$ \\
PaCMAP     & $0.044\pm0.028$ & $0.160\pm0.020$ & $0.874\pm0.006$ & $0.900\pm0.021$ \\
PHATE      & $0.269\pm0.033$ & $0.338\pm0.035$ & $0.896\pm0.035$ & $0.754\pm0.034$ \\
DMT-Dens   & \best{$0.811\pm0.011$} & \best{$0.880\pm0.008$} & $0.853\pm0.043$ & $0.846\pm0.038$ \\
\botrule
\end{tabular*}
\end{table}

\begin{figure}[H]
\centering
\includegraphics[width=\columnwidth]{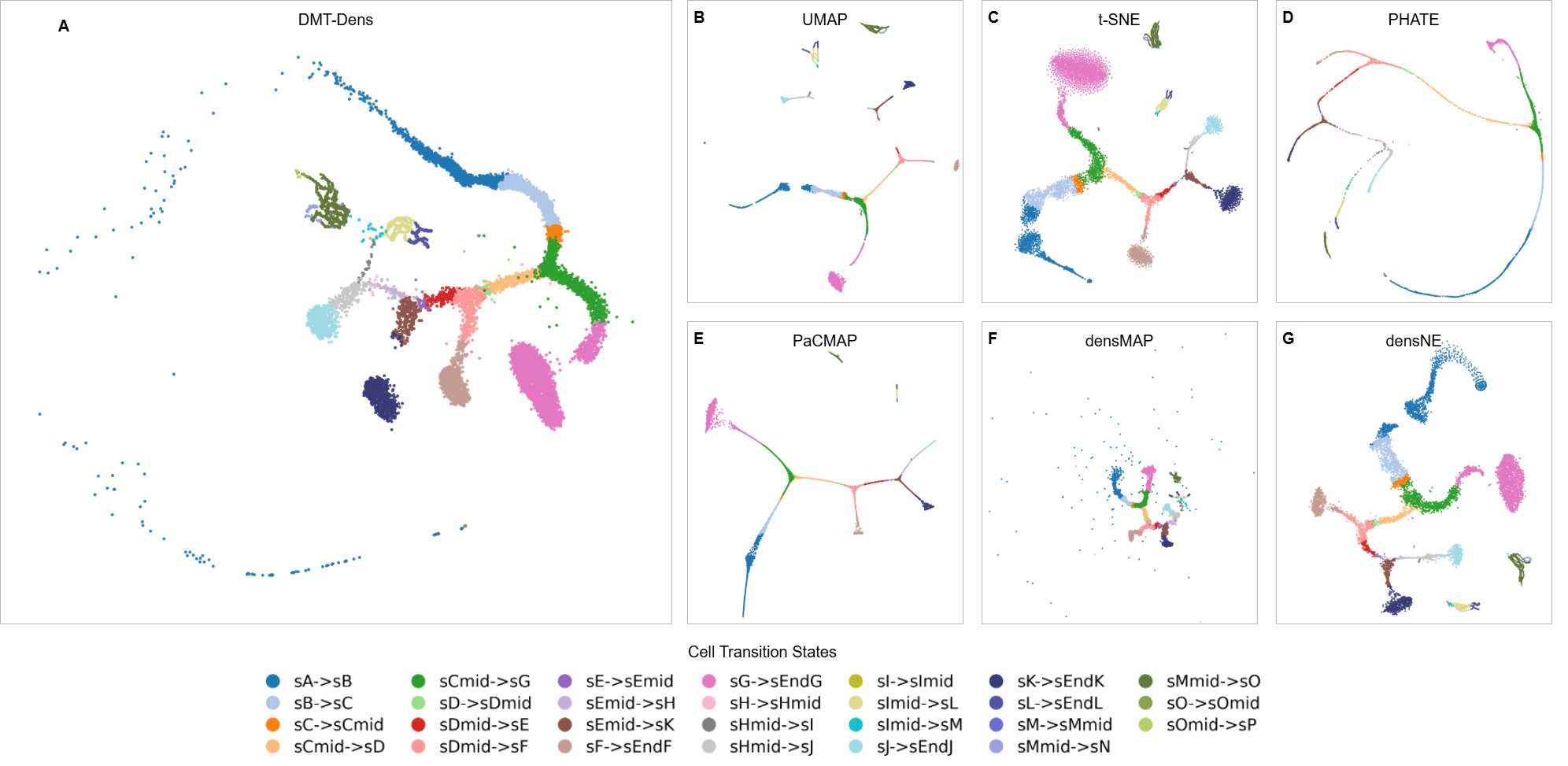}
\caption{Embeddings of synthetic \texttt{dyngen} data with a known branching topology. Colors denote the reference branch assignment.\label{fig:supp-dyngen}}
\end{figure}

Across five seeds, DMT-Dens records the highest density correlation ($0.811\pm0.011$) and local density correlation ($0.880\pm0.008$), compared with $0.754\pm0.009$ and $0.801\pm0.003$, respectively, for densMAP. Its Branch SVC is $0.853\pm0.043$, while t-SNE records the highest value of $0.922\pm0.003$. Its topology fidelity is $0.846\pm0.038$, while den-SNE records the highest value of $0.923\pm0.000$. Figure~\ref{fig:supp-dyngen} shows the embeddings colored by reference branch assignment.

\subsection{Developmental-time preservation (\emph{C.\ elegans})}\label{sec:supp-celegan-traj}
We next assess how each embedding represents developmental time in the \emph{C.\ elegans} embryonic time course \citep{packer2019celegan}. Each cell has a binned embryonic-time annotation, which we map to a numeric value. Because the dataset does not provide a lineage backbone, the evaluation is based on the observed time annotation and the two-dimensional embedding.

\emph{Pseudotime correlation} is the Spearman correlation between observed embryonic time and embedding pseudotime. Embedding pseudotime is defined as geodesic distance from the earliest-time cells on the embedding $k$-nearest-neighbor graph ($k=15$). \emph{Ordering accuracy} is the tie-aware fraction of cell pairs with different observed times whose embedding-pseudotime order agrees with their observed-time order. \emph{Time continuity} is one minus the mean absolute observed-time difference across embedding kNN edges, normalized by the corresponding mean for random pairs; higher values indicate greater temporal similarity between neighboring cells. \emph{DEMaP} is the Spearman correlation between ambient-space geodesic distances and embedding-space Euclidean distances.

We also report \emph{reach}, the fraction of cells reachable from the earliest-time root on the embedding graph. For fragmented graphs, pseudotime correlation and ordering accuracy are computed only on the reachable subset and must therefore be interpreted together with reach. Entries are mean$\pm$sample standard deviation over five seeds. Unmarked den-SNE and densMAP rows use their default density weights (0.1 and 2.0, respectively); rows marked ``$(\lambda)$'' use the density weight selected in the second-stage search while holding the structural hyperparameters fixed.

\begin{table}[H]
\caption{Developmental-time preservation relative to observed embryonic-time annotations in the \emph{C.\ elegans} time course. Entries are mean$\pm$sample s.d.\ over five seeds; higher values are better. For fragmented embeddings, pseudotime correlation and ordering accuracy are computed on the reachable subset and must be interpreted together with reach. No rank-based highlighting is applied. Rows marked ``$(\lambda)$'' use the density weight selected in the second-stage search; unmarked rows use the default weight.\label{tab:supp-celegan-traj}}
\centering
\scriptsize
\setlength{\tabcolsep}{3pt}
\renewcommand{\arraystretch}{1.04}
\begin{tabular*}{\columnwidth}{@{\extracolsep{\fill}}lccccc@{}}
\toprule
Method & Reach & Pseudotime corr. & Ordering acc. & Time continuity & DEMaP \\
\midrule
t-SNE      & $0.921\pm0.011$ & $0.285\pm0.089$ & $0.615\pm0.036$ & $0.776\pm0.001$ & $0.428\pm0.018$ \\
UMAP       & $0.630\pm0.043$ & $-0.133\pm0.113$ & $0.446\pm0.046$ & $0.716\pm0.002$ & $0.354\pm0.020$ \\
den-SNE    & $0.624\pm0.000$ & $0.184\pm0.000$ & $0.568\pm0.000$ & $0.771\pm0.000$ & $0.484\pm0.000$ \\
densMAP    & $1.000\pm0.000$ & $0.212\pm0.048$ & $0.585\pm0.019$ & $0.448\pm0.005$ & $0.565\pm0.013$ \\
PaCMAP     & $0.430\pm0.085$ & $0.518\pm0.089$ & $0.713\pm0.042$ & $0.783\pm0.003$ & $0.502\pm0.007$ \\
PHATE      & $1.000\pm0.000$ & $0.392\pm0.050$ & $0.657\pm0.020$ & $0.557\pm0.012$ & $0.817\pm0.014$ \\
den-SNE ($\lambda$) & $1.000\pm0.000$ & $0.234\pm0.000$ & $0.593\pm0.000$ & $0.649\pm0.000$ & $0.629\pm0.000$ \\
densMAP ($\lambda$) & $1.000\pm0.000$ & $0.263\pm0.049$ & $0.604\pm0.020$ & $0.401\pm0.006$ & $0.615\pm0.005$ \\
DMT-Dens & $0.997\pm0.001$ & $0.556\pm0.052$ & $0.732\pm0.027$ & $0.739\pm0.005$ & $0.643\pm0.011$ \\
\botrule
\end{tabular*}
\end{table}

DMT-Dens obtains a reach of $0.997\pm0.001$, a pseudotime correlation of $0.556\pm0.052$, and an ordering accuracy of $0.732\pm0.027$. The corresponding PaCMAP values are $0.430\pm0.085$, $0.518\pm0.089$, and $0.713\pm0.042$, respectively. Time continuity is $0.739\pm0.005$ for DMT-Dens, $0.776\pm0.001$ for t-SNE, and $0.783\pm0.003$ for PaCMAP. DEMaP is $0.643\pm0.011$ for DMT-Dens and $0.817\pm0.014$ for PHATE.

\section{Full embedding comparisons}
\begin{figure}[H]
\centering
\includegraphics[width=\textwidth,height=0.75\textheight,keepaspectratio]{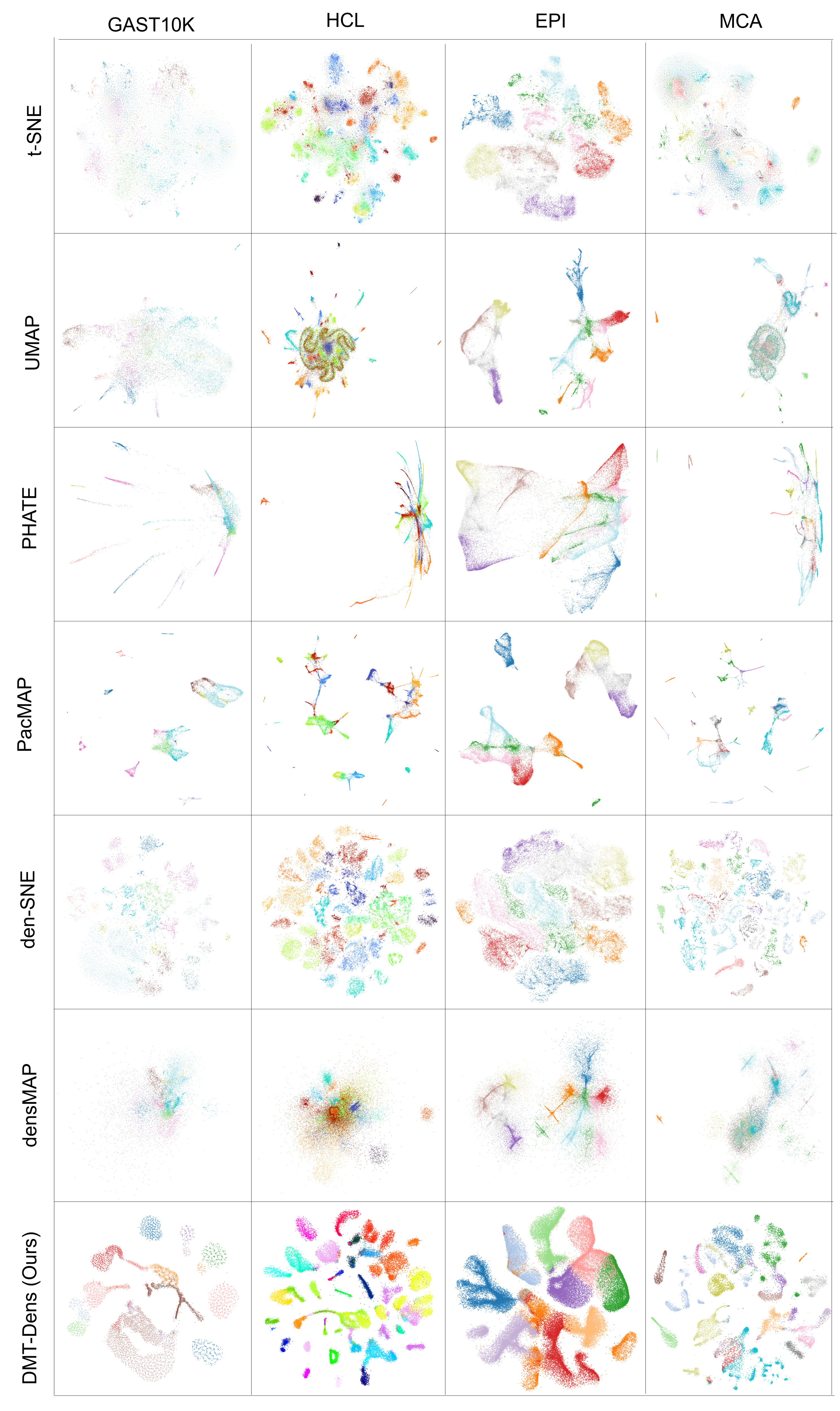}
\caption{Full embedding comparison across seven methods on the four single-cell datasets (GAST10K, HCL, EPI, and MCA). Points are colored by cell-type label. This figure extends main-text Figure~\ref{fig:comparison}.\label{fig:comparison-full}}
\end{figure}

\begin{figure}[H]
\centering
\includegraphics[width=\textwidth,height=0.75\textheight,keepaspectratio]{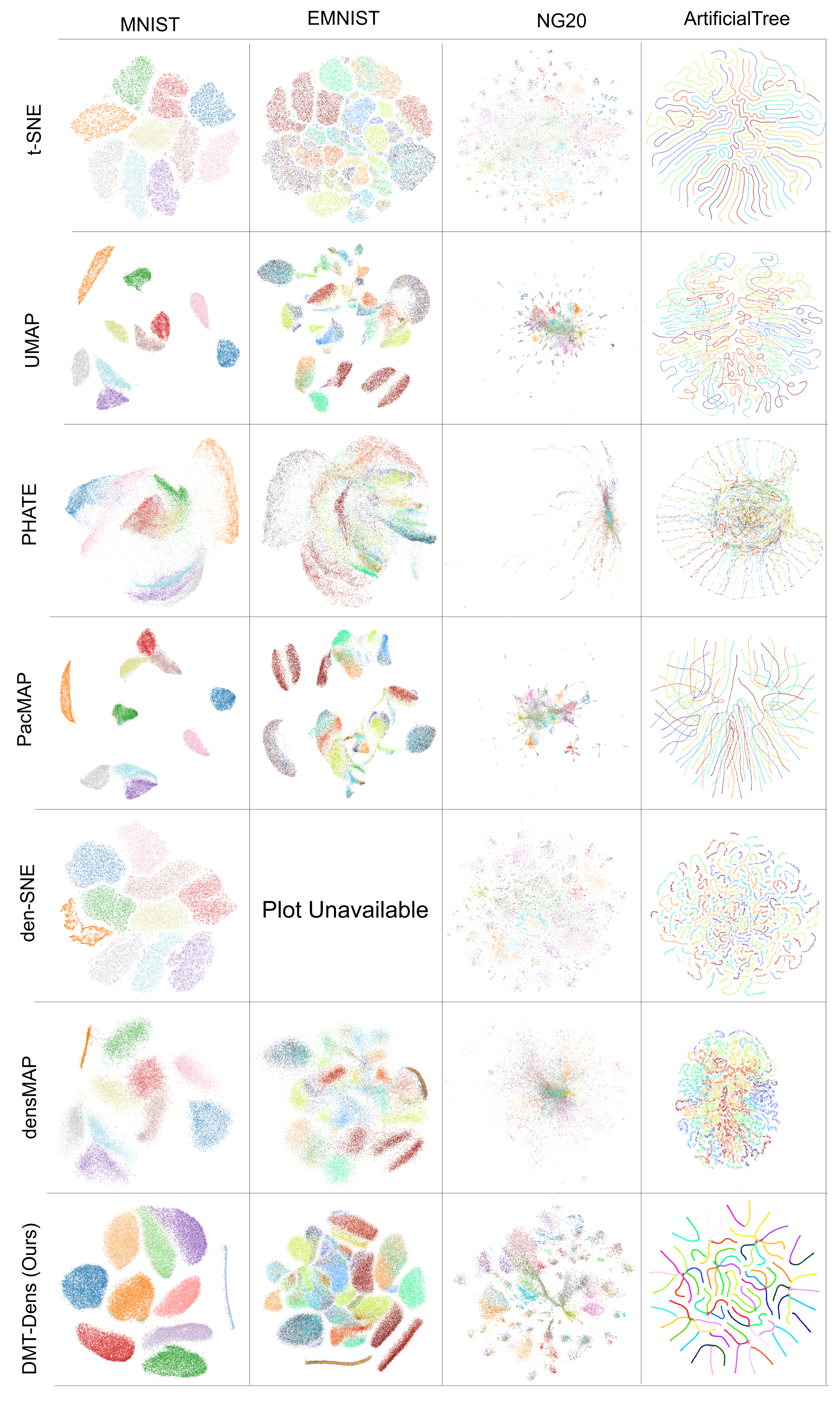}
\caption{Full embedding comparison across seven methods on MNIST, EMNIST, 20~Newsgroups, and ArtificialTree. Points are colored by class label. The panels illustrate qualitative differences in density variation, overlap, and fragmentation; quantitative density and label-separability results are reported separately.\label{fig:text-comparison-full}}
\end{figure}

Figures~\ref{fig:comparison-full} and~\ref{fig:text-comparison-full} extend the representative embedding comparison in main-text Figure~\ref{fig:comparison} to all seven methods. Dataset columns, color scales, and point styling match those in the main text.

%% file: topobranch_refs.bib
@article{maaten2008visualizing,
  title={Visualizing data using t-{SNE}},
  author={van der Maaten, Laurens and Hinton, Geoffrey},
  journal={Journal of Machine Learning Research},
  volume={9},
  number={86},
  pages={2579--2605},
  year={2008},
  url={https://www.jmlr.org/papers/v9/vandermaaten08a.html}
}

@article{mcinnes2018umap,
  title={{UMAP}: Uniform manifold approximation and projection for dimension reduction},
  author={McInnes, Leland and Healy, John and Melville, James},
  journal={arXiv preprint arXiv:1802.03426},
  year={2018},
  doi={10.48550/arXiv.1802.03426}
}

@article{becht2019umap,
  title={Dimensionality reduction for visualizing single-cell data using {UMAP}},
  author={Becht, Etienne and McInnes, Leland and Healy, John and Dutertre, Charles-Antoine and Kwok, Immanuel W H and Ng, Lai Guan and Ginhoux, Florent and Newell, Evan W},
  journal={Nature Biotechnology},
  volume={37},
  number={1},
  pages={38--44},
  year={2019},
  doi={10.1038/nbt.4314}
}

@article{moon2019phate,
  title={Visualizing structure and transitions in high-dimensional biological data},
  author={Moon, Kevin R and van Dijk, David and Wang, Zheng and Burkhardt, Daniel B and Chen, William S and van den Elzen, Antal and Hirn, Matthew J and Coifman, Ronald R and Ivanova, Natalia B and Wolf, Guy and Krishnaswamy, Smita},
  journal={Nature Biotechnology},
  volume={37},
  number={12},
  pages={1482--1492},
  year={2019},
  doi={10.1038/s41587-019-0336-3}
}

@article{narayan2021assessing,
  title={Assessing single-cell transcriptomic variability through density-preserving data visualization},
  author={Narayan, Ashwin and Berger, Bonnie and Cho, Hyunghoon},
  journal={Nature Biotechnology},
  volume={39},
  number={6},
  pages={765--774},
  year={2021},
  doi={10.1038/s41587-020-00801-7}
}

@article{wolf2019paga,
  title={{PAGA}: graph abstraction reconciles clustering with trajectory inference through a topology preserving map of single cells},
  author={Wolf, F Alexander and Hamey, Fiona K and Plass, Mireya and Solana, Jordi and Dahlin, Joakim S and G{\"o}ttgens, Berthold and Rajewsky, Nikolaus and Simon, Lukas and Theis, Fabian J},
  journal={Genome Biology},
  volume={20},
  number={1},
  pages={59},
  year={2019},
  doi={10.1186/s13059-019-1663-x}
}

@article{setty2019palantir,
  title={Characterization of cell fate probabilities in single-cell data with {Palantir}},
  author={Setty, Manu and Kiseliovas, Vaidotas and Levine, Jacob and Gayoso, Adam and Mazutis, Linas and Pe'er, Dana},
  journal={Nature Biotechnology},
  volume={37},
  number={4},
  pages={451--460},
  year={2019},
  doi={10.1038/s41587-019-0068-4}
}

@article{saelens2019comparison,
  title={A comparison of single-cell trajectory inference methods},
  author={Saelens, Wouter and Cannoodt, Robrecht and Todorov, Helena and Saeys, Yvan},
  journal={Nature Biotechnology},
  volume={37},
  number={5},
  pages={547--554},
  year={2019},
  doi={10.1038/s41587-019-0071-9}
}

@article{wang2021pacmap,
  title={Understanding how dimension reduction tools work: An empirical approach to deciphering t-{SNE}, {UMAP}, {TriMap}, and {PaCMAP} for data visualization},
  author={Wang, Yingfan and Huang, Haiyang and Rudin, Cynthia and Shaposhnik, Yaron},
  journal={Journal of Machine Learning Research},
  volume={22},
  number={201},
  pages={1--73},
  year={2021},
  url={https://www.jmlr.org/papers/v22/20-1061.html}
}

@article{bunne2023cellot,
  title={Learning single-cell perturbation responses using neural optimal transport},
  author={Bunne, Charlotte and Stark, Stefan G and Gut, Gabriele and {Sarabia del Castillo}, Jacobo and Levesque, Mitch and Lehmann, Kjong-Van and Pelkmans, Lucas and Krause, Andreas and R{\"a}tsch, Gunnar},
  journal={Nature Methods},
  volume={20},
  number={11},
  pages={1759--1768},
  year={2023},
  doi={10.1038/s41592-023-01969-x}
}

@article{biondo2025intrinsic,
  title={The intrinsic dimension of gene expression during cell differentiation},
  author={Biondo, Marta and Cirone, Niccol{\`o} and Valle, Filippo and Lazzardi, Silvia and Caselle, Michele and Osella, Matteo},
  journal={Nucleic Acids Research},
  volume={53},
  number={16},
  pages={gkaf805},
  year={2025},
  doi={10.1093/nar/gkaf805}
}

@article{otto2024mellon,
  title={Quantifying cell-state densities in single-cell phenotypic landscapes using {Mellon}},
  author={Otto, Dominik Jenz and Jordan, Cailin and Dury, Brennan and Dien, Christine and Setty, Manu},
  journal={Nature Methods},
  volume={21},
  number={7},
  pages={1185--1195},
  year={2024},
  doi={10.1038/s41592-024-02302-w}
}

@article{vandenbrink2017single,
  title={Single-cell sequencing reveals dissociation-induced gene expression in tissue subpopulations},
  author={{van den Brink}, Susanne C. and Sage, Fanny and V{\'e}rtesy, {\'A}bel and Spanjaard, Bastiaan and Peterson-Maduro, Josi and Baron, Chlo{\'e} S. and Robin, Catherine and {van Oudenaarden}, Alexander},
  journal={Nature Methods},
  volume={14},
  number={10},
  pages={935--936},
  year={2017},
  doi={10.1038/nmeth.4437}
}

@article{bergen2020velocity,
  title={Generalizing RNA velocity to transient cell states through dynamical modeling},
  author={Bergen, Volker and Lange, Marius and Peidli, Stefan and Wolf, F Alexander and Theis, Fabian J},
  journal={Nature Biotechnology},
  volume={38},
  number={12},
  pages={1408--1414},
  year={2020},
  doi={10.1038/s41587-020-0591-3}
}

@article{lange2022cellrank,
  title={CellRank for directed single-cell fate mapping},
  author={Lange, Marius and Bergen, Volker and Klein, Michal and Setty, Manu and Reuter, Beate and Bakhti, Mostafa and L{\"u}ck, Stefan and Ansari, Mehrnoush and Schniering, Jens and Schiller, Herbert B and Peidli, Stefan and Wolf, F Alexander and Theis, Fabian J},
  journal={Nature Methods},
  volume={19},
  number={2},
  pages={159--170},
  year={2022},
  doi={10.1038/s41592-021-01346-6}
}

@article{kobak2019art,
  title={The art of using t-SNE for single-cell transcriptomics},
  author={Kobak, Dmitry and Berens, Philipp},
  journal={Nature Communications},
  volume={10},
  number={1},
  pages={5416},
  year={2019},
  doi={10.1038/s41467-019-13056-x}
}

@article{kobak2021initialization,
  title={Initialization is critical for preserving global data structure in both t-{SNE} and {UMAP}},
  author={Kobak, Dmitry and Linderman, George C},
  journal={Nature Biotechnology},
  volume={39},
  number={2},
  pages={156--157},
  year={2021},
  doi={10.1038/s41587-020-00809-z}
}

@inproceedings{nguyen2019diffusion,
  title={Diffusion t-SNE for multiscale data visualization},
  author={Nguyen, Lan Huong and Holmes, Susan},
  booktitle={Proceedings of the Machine Learning for Computational Biology Conference},
  year={2019},
  url={https://mlcb.github.io/mlcb2019_proceedings/papers/paper_45.pdf}
}

@article{chari2023specious,
  title={The specious art of single-cell genomics},
  author={Chari, Tara and Pachter, Lior},
  journal={PLOS Computational Biology},
  volume={19},
  number={8},
  pages={e1011288},
  year={2023},
  doi={10.1371/journal.pcbi.1011288}
}

@article{xu2023dmt,
  title={Structure-preserving visualization for single-cell {RNA}-Seq profiles using deep manifold transformation with batch-correction},
  author={Xu, Yongjie and Zang, Zelin and Xia, Jun and Tan, Cheng and Geng, Yulan and Li, Stan Z.},
  journal={Communications Biology},
  volume={6},
  number={1},
  pages={369},
  year={2023},
  doi={10.1038/s42003-023-04662-z}
}

@article{xu2025poincaredmt,
  title={Complex hierarchical structures analysis in single-cell data with {Poincar{\'e}} deep manifold transformation},
  author={Xu, Yongjie and Zang, Zelin and Hu, Bozhen and Yuan, Yue and Tan, Cheng and Xia, Jun and Li, Stan Z.},
  journal={Briefings in Bioinformatics},
  volume={26},
  number={1},
  pages={bbae687},
  year={2025},
  doi={10.1093/bib/bbae687}
}

@article{zang2025must,
  title={{MuST}: multiple-modality structure transformation for single-cell spatial transcriptomics},
  author={Zang, Zelin and Li, Liangyu and Xu, Yongjie and Duan, Chenrui and Shen, Yue and Sun, Yi and Lei, Zhen and Li, Stan Z.},
  journal={Briefings in Bioinformatics},
  volume={26},
  number={4},
  pages={bbaf405},
  year={2025},
  doi={10.1093/bib/bbaf405}
}

@article{zhou2021dissecting,
  title={Dissecting transition cells from single-cell transcriptome data through multiscale stochastic dynamics},
  author={Zhou, Peijie and Wang, Shuxiong and Li, Tiejun and Nie, Qing},
  journal={Nature Communications},
  volume={12},
  number={1},
  pages={5609},
  year={2021},
  doi={10.1038/s41467-021-25548-w}
}

@article{cannoodt2021dyngen,
  title={Spearheading future omics analyses using dyngen, a multi-modal simulator of single cells},
  author={Cannoodt, Robrecht and Saelens, Wouter and Deconinck, Louise and Saeys, Yvan},
  journal={Nature Communications},
  volume={12},
  number={1},
  pages={3942},
  year={2021},
  doi={10.1038/s41467-021-24152-2}
}

@article{han2018mca,
  title={Mapping the mouse cell atlas by {Microwell-Seq}},
  author={Han, Xiaoping and Wang, Renying and Zhou, Yincong and Fei, Lijiang and Sun, Huiyu and Lai, Shujing and Saadatpour, Assieh and Zhou, Ziming and Chen, Haide and Ye, Fang and Huang, Daosheng and Xu, Yang and Huang, Wentao and Jiang, Mengmeng and Jiang, Xinyi and Mao, Jie and Chen, Yao and Lu, Chenyu and Xie, Jin and Fang, Qun and Wang, Yibin and Yue, Rui and Li, Tiefeng and Huang, He and Orkin, Stuart H and Yuan, Guo-Cheng and Chen, Ming and Guo, Guoji},
  journal={Cell},
  volume={172},
  number={5},
  pages={1091--1107},
  year={2018},
  doi={10.1016/j.cell.2018.02.001}
}

@article{han2020hcl,
  title={Construction of a human cell landscape at single-cell level},
  author={Han, Xiaoping and Zhou, Ziming and Fei, Lijiang and Sun, Huiyu and Wang, Renying and Chen, Yao and Chen, Haide and Wang, Jingjing and Tang, Huanna and Ge, Wenhao and Zhou, Yincong and Ye, Fang and Jiang, Mengmeng and Wu, Junqing and Xiao, Yanyu and Jia, Xiaoning and Zhang, Tingyue and Ma, Xiajun and Zhang, Qi and Bai, Xueli and Lai, Shujing and Yu, Chengxuan and Zhu, Lijun and Lin, Rui and Gao, Yuchi and Wang, Min and Wu, Yiqing and Zhang, Jianming and Zhan, Renya and Zhu, Saiyong and Hu, Hailan and Wang, Changchun and Chen, Ming and Huang, He and Liang, Tingbo and Chen, Jianghua and Wang, Weilin and Zhang, Dan and Guo, Guoji},
  journal={Nature},
  volume={581},
  number={7808},
  pages={303--309},
  year={2020},
  doi={10.1038/s41586-020-2157-4}
}

@article{packer2019celegan,
  title={A lineage-resolved molecular atlas of {C. elegans} embryogenesis at single-cell resolution},
  author={Packer, Jonathan S and Zhu, Qin and Huynh, Chau and Sivaramakrishnan, Priya and Preston, Elicia and Dueck, Hannah and Stefanik, Derek and Tan, Kai and Trapnell, Cole and Kim, Junhyong and Waterston, Robert H and Murray, John I},
  journal={Science},
  volume={365},
  number={6459},
  pages={eaax1971},
  year={2019},
  doi={10.1126/science.aax1971}
}

@article{zhang2019gastric,
  title={Dissecting the single-cell transcriptome network underlying gastric premalignant lesions and early gastric cancer},
  author={Zhang, Peng and Yang, Minghui and Zhang, Yuan and Xiao, Shengnan and Lai, Xinxin and Tan, Aiping and Du, Shaojun and Li, Shanshan},
  journal={Cell Reports},
  volume={27},
  number={6},
  pages={1934--1947.e5},
  year={2019},
  doi={10.1016/j.celrep.2019.04.052}
}

@inproceedings{vaswani2017attention,
  author    = {Vaswani, Ashish and Shazeer, Noam and Parmar, Niki and Uszkoreit, Jakob and Jones, Llion and Gomez, Aidan N. and Kaiser, {\L}ukasz and Polosukhin, Illia},
  title     = {Attention Is All You Need},
  booktitle = {Advances in Neural Information Processing Systems},
  volume    = {30},
  year      = {2017},
  url       = {https://papers.neurips.cc/paper_files/paper/2017/hash/3f5ee243547dee91fbd053c1c4a845aa-Abstract.html}
}
